\documentclass[10pt,aps,pra,twocolumn,showpacs,superscriptaddress]{revtex4-1}
\usepackage[english]{babel} 
\usepackage[usenames, dvipsnames]{color} 
\usepackage{graphicx} 
\usepackage{bm} 
\usepackage{amsmath} 
\usepackage{amsthm} 
\usepackage{amssymb} 
\usepackage{times} 
\usepackage[pdftex,colorlinks=true,urlcolor= blue,linkcolor=Blue,citecolor=RedViolet]{hyperref}
\usepackage{hyperref}

\DeclareMathOperator{\arccosh}{arccosh}

\begin{document}
\title{Transport of Correlations in a Harmonic Chain}
\author{F. Nicacio} \email{fernando.nicacio@ufabc.edu.br  } 
\author{F. L. Semi\~ao}
\affiliation{ Centro de Ci\^encias Naturais e Humanas, 
              Universidade Federal do ABC,   
              09210-170, Santo Andr\'e, S\~ao Paulo, Brazil }
\date{\today}
\begin{abstract}
\noindent We study the propagation of different types of 
correlations through a quantum bus formed by a chain of coupled harmonic oscillators. 
This includes steering, entanglement, mutual information, quantum discord and 
Bell-like nonlocality. 
The  whole system consists of the quantum bus (propagation medium) and other quantum 
harmonic oscillators (sources and receivers of quantum correlations) weakly coupled 
to the chain. We are particularly interested in using the point of view of transport to 
spot distinctive features displayed by different kinds of correlations. 
We found, for instance, that there are fundamental differences in the way steering 
and discord propagate, depending on the way they are defined with respect to the parties 
involved in the initial correlated  state. 
We analyzed both the closed and open system dynamics as well as the role played by 
thermal excitations in the propagation of the correlations.  
\end{abstract}
\maketitle
\section{ Introduction }\label{introd}  
A direct consequence of the inherent probabilistic nature of measurement outputs and 
state superposition of the quantum world is the 
existence of correlations that can not be explained through a classical description. 
Although most of these correlations have been well studied 
\cite{gisin,horodeckis,amico,jones,ollivier,henderson,
      groisman,wiseman},  
their features in the problem of transport are yet not fully explored in the literature, 
except for a few exceptional cases \cite{plenio2004,plenio2005}. 
It is still missing, however, 
a comparative study dedicated to contrast the way different correlations 
are transported in a chain of coupled quantum systems. 
The present work is dedicated to this matter. 
This kind of study recognizes that a fundamental requisite to perform quantum 
communication is precisely the ability to move information/correlations around quantum 
systems, and tries to use this paradigm to compare different correlations as 
propagation is involved.

The propagation of entanglement through harmonic networks
is one of the few examples which are well explored in the literature 
in \cite{plenio2004,plenio2005}. 
Notably, in \cite{plenio2005} a useful scheme for high efficiency entanglement 
propagation, under unitary dynamics (closed system), is developed for a linear chain. 
The proposed method consists of minimal adjustments of frequencies and 
coupling constants to boost the efficiency. The result is that the chain 
is transformed in a quantum bus able to distribute entanglement among quantum systems 
coupled to it. To understand this result, a simple yet accurate model obtained 
using the rotating wave approximation (RWA) is employed.  
In \cite{nicacio4}, symplectic tools are used to generalize this approach and to arrive at 
an effective and accurate simple description suitable for arbitrary topologies 
(beyond linear chains) and open system dynamics (interaction with 
environment).  

The main goal of our present work is to apply the system described in \cite{plenio2005} 
to the study of transport of different quantum correlations, not only entanglement. 
We are interested, for instance, in how an intrinsically asymmetric correlation such as 
quantum steering propagates. It is well known that given a state of a composed system 
$A\otimes B$, steering $AB$ may be different than steering $BA$ \cite{jones,wiseman}. 
How does this difference show up in a propagation problem? For our knowledge, 
it is the first time steering is studied under this perspective. 
Additionally, we do not restrict our investigation to entanglement and steering as we 
also include propagation of discord \cite{ollivier,henderson}, 
mutual information \cite{henderson,groisman} 
and Bell-like nonlocality \cite{jones,wiseman,gisin} 
to end up with a broad view of the propagations of quantum correlations using a quantum bus. 
We also include the action of thermal baths to mimic the effects of losses during the 
transmission of the correlations. For that, we apply the method developed 
in \cite{nicacio4} to arrive at an effective description of the system dynamics which 
allowed us to obtain physically relevant analytical expressions to explain the main 
features of the propagation problem. 
Numerical solutions of the exact equations of motion are also always employed to support 
the validity of the analytical treatment.

The question of how such different types of correlation propagate in a chain of coupled 
quantum systems is by itself a question of theoretical relevance. 
However, it goes beyond that since moving correlations around  
in large systems seems to be relevant to the successful use of quantum information in 
practical terms.
Today, we see an unprecedented level of control over nano- and mesoscopic systems 
where such an ability to propagate correlations may be required in near future. 
In this scope, the harmonic movement of interacting nanoelectromechanical 
oscillators \cite{buks}  
and systems consisting of dipole-dipole interacting trapped ions \cite{bermudez} 
are illustrative examples. 

This article is organized as follows. 
In Sec.~{\ref{tsid}}, we describe the physical setup used in this work. 
This includes the quantum bus \cite{plenio2005}, used to establish correlations among 
external oscillators attached to it, and the procedure put forward in \cite{nicacio4} to treat 
the open system case. 
In Sec.~ {\ref{QC}}, we review the concepts that motivate the definition of the different 
correlations used in this work and particularize them to the bipartite Gaussian-state 
scenario.
The propagation phenomenon itself and the comparison between entanglement and steering are 
given in Sec.\ref{TQC}, while Sec.~\ref{miaqd} discussed discord and mutual information. 
Section \ref{pobc} is devoted to the propagation of Bell-like nonlocality. 
We end the work with final remarks summarizing the main results and discussing possible 
generalizations. 

\section{ The System and Its Dynamics }\label{tsid}  
\begin{figure}[htb!]                                                      
\includegraphics[width=8cm]{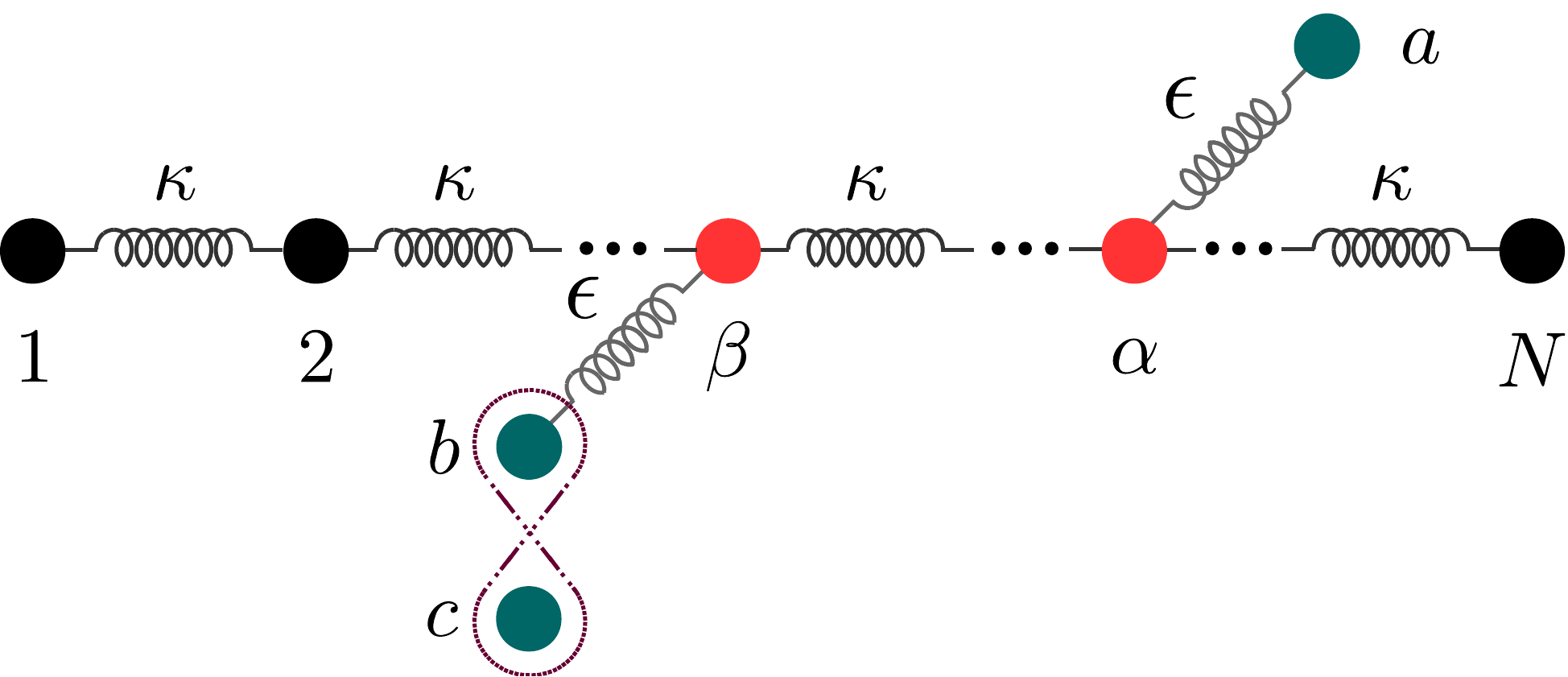} 
\caption{ (Color Online) Schematic representation of the system. 
It consists of a chain with $N$-coupled harmonic oscillators, 
where the $\alpha^\text{\underline{th}}$ and 
$\beta^\text{\underline{th}}$ oscillators are also coupled to 
oscillators $a$ and $b$, respectively. 
Coupling constants $\kappa$ and $\epsilon $ refer to Hooke-type interactions. 
Oscillator $c$ is decoupled from the chain but initially correlated 
to oscillator $b$. Thermal baths (not shown) are coupled to each oscillator.
}                                                                                        \label{fig1system1}                            
\end{figure}

The physical system considered in this work is depicted in Fig.~\ref{fig1system1}. 
Our aim is to describe the transfer of correlations, originally present in the state of 
noninteracting oscillators $c$ and $b$, to oscillators $c$ and $a$ throughout the 
$N$-oscillator chain. 
Notice that $c$ is totally decoupled from the chain, such that the latter works 
really as just a distributor of the part of correlation which comes originally from $b$. 

The temporal evolution of the global system will be governed by a Lindblad master 
equation (LME) for the density operator $\hat \rho$:
\begin{equation}                                                                         \label{lindblad}
\frac{d \hat \rho }{d t}  = 
         \frac{i}{\hbar}  [ \hat \rho , \hat H ] + \mathcal L(\hat\rho), 
\end{equation} 
where $\hat H$ is the Hamiltonian of the system and $\mathcal{L}$ is 
the Lindblad generator accounting for the environment-system interaction. 
For the chain in Fig.~\ref{fig1system1}, we write 
$ \hat H = \hat H_0 + \hat H_I$, and also $\hat H_0 = \hat H_{N} + \hat H_{\rm E}$, 
where
\begin{equation}                                                                         \label{hamchain}
\hat H_{N} = 
\frac{\omega}{2} \sum_{k = 1}^{N}  (\hat q_k^2 + \hat p_k^2) +
\frac{\kappa}{4} \sum_{k = 1}^{N-1}(\hat q_{k+1} - \hat q_k)^2
\end{equation}
is the Hamiltonian of the $N$-oscillator chain,   
\begin{equation}                                                                         \label{hamext}
\hat H_{\rm E} = \sum_{k = a,b,c} \frac{\Omega_k}{2}
                  (\hat {\sf q}_k^2 + \hat {\sf p}_k^2) 
\end{equation}
is the free Hamiltonian of the external oscillators, 
and the interaction with the chain is governed by
\begin{equation}
\hat H_{\rm I} = \frac{\epsilon}{4} (\hat q_{\alpha} - \hat {\sf q}_a)^2 + 
             \frac{\epsilon}{4} (\hat q_{\beta} - \hat {\sf q}_b)^2 .
\end{equation}

The nonunitary part of the dynamics, resulting from the unavoidable 
incapacity of perfect system isolation, is modeled here through a 
Lindblad generator originating from interaction with local thermal baths, {\it i.e.}, 
\begin{equation}                                                                         \label{lopthermal}     
\begin{aligned} 
\mathcal L (\hat \rho) =& - \frac{1}{2}\zeta ({\bar n}_{\rm th} + 1)
\sum_{k} \!\! 
\left( 
\{ \hat{A}_k^\dag \hat{A}_k , \hat \rho  \}
- 2 \hat{A}_k \hat \rho \hat{A}_k^\dag   \right)     \\
&- \frac{1}{2}\zeta \,{\bar n}_{\rm th}
\sum_{k} \!\! 
\left( 
\{ \hat{A}_k \hat{A}_k^\dag , \hat \rho  \}
- 2 \hat{A}_k^\dag \hat \rho \hat{A}_k   \right), 
\end{aligned}
\end{equation}
%
where 
$\hat A_k := (\hat q_k + i \hat p_k)/\sqrt{2\hbar}$ $(k = a,b,c,1,2,...,N)$
are the annihilation operators, 
$\zeta \ge 0$ is a bath-oscillator coupling constant, and 
${\bar n}_{\rm th}$ is a thermal occupation number. 
These two parameters are taken to be the same for all oscillators 
without loss of generality for the purposes of our work.  

As stated before, we want to investigate how quantum correlations, originally present in 
the state of noninteracting oscillators $c$ and $b$, are redistributed by the 
$N$-oscillator chain to oscillators $c$ and $a$. 
These correlations will be retrieved from state of the subsystem $[ac]$, obtained from 
the global density matrix $\hat \rho(t)$ by tracing out the remaining $N+1$ oscillators.
Since the Hamiltonian and the Lindblad operators are, respectively, quadratic and linear 
in the position and momentum operators, initial Gaussian states will evolve to Gaussian 
states. 
In this case, there is a general recipe which allows one to write a formal solution for 
the evolved covariance matrix  \cite{nicacio4, nicacio}, and from that, the density 
matrix $\hat \rho(t)$ can be obtained. 
However, in practice, the story is a bit more intricate. 
Despite the well-known form of the solution, its use requires the exponentiation of 
$(2N + 6) \times (2N + 6)$ matrices \cite{nicacio4}. 
This becomes impractical as soon as the number of oscillators in the chain becomes 
moderately high. 
Having this in mind, one is usually driven to employ methods that effectively reduce the 
size of the problem such as the one developed in \cite{nicacio4}. 
In what follows, we give a brief description of this method applied to the system 
depicted in Fig.~\ref{fig1system1}. 

The starting point is to find the eigenfrequencies of the Hamiltonian $\hat H_N$ 
in (\ref{hamchain}), denoted here as $\varsigma_k$ for $k = 1,...,N$. 
By setting the frequencies of the external oscillators $\Omega_k$ in (\ref{hamext}) 
equal to any of the eigenfrequencies of the chain, for instance, 
by taking $\Omega_{a,b,c} = \varsigma_m$  for a fixed $m$ ($1\leq m\leq N$), 
a RWA is applied to (\ref{lindblad}), leading to a simplified Lindblad equation 
involving just the subsystem  $(a,b,c,m)$. 
This will be an accurate description as long as  the external oscillators are 
weakly coupled to the chain $\epsilon \ll \kappa,\varsigma_k$. 
The effective density matrix for this reduced problem will be called $\check \rho$,
and its evolution in the interaction picture, with respect to $\hat H_0$, is governed by 
\begin{equation}                                                                         \label{lindeff}
\frac{d \check \rho }{d t}  = 
         \frac{i}{\hbar}  [ \check \rho , \check H ] + \mathcal{\check L}(\check \rho)
         + \mathcal{\check L}'(\check \rho), 
\end{equation} 
with the effective Hamiltonian
\begin{eqnarray}                                                                          \label{hameff} 
&&\check H  = 
\frac{\hbar\epsilon\omega}{4 { \varsigma_m }} 
\left(  \pmb{O}_{\! m \alpha}^{2}  +  \pmb{O}_{\! m \beta }^{2} \right) 
\hat a_m^\dagger\hat a_m  +
\frac{\hbar\epsilon}{4} ( \hat A^\dagger_a \hat A_a +  \hat A^\dagger_b \hat A_b   )  \\
&&- \frac{\hbar\epsilon\sqrt{\omega}  }{4 \sqrt{ \varsigma_m} } 
\left[ 
\pmb{O}_{ \! m \alpha }( \hat a_m \hat A_a^{\dag} + \hat a_m^{\dag} \hat A_a  )  + 
\pmb{O}_{ \! m \beta } ( \hat a_m \hat A_b^{\dag} + \hat a_m^{\dag} \hat A_b  ) 
\right],                                                                                 \nonumber
\end{eqnarray}
where $\hat A_k := (\hat{\sf q}_k + i \hat {\sf p}_k)/\sqrt{2\hbar} \,\, (k = a,b,c$)
are the annihilation operators for the external oscillators,   
$\hat a_m:= (\hat{Q}_m + i \hat {P}_m)/\sqrt{2\hbar}$ 
is the annihilation operator associated with the position and momentum operators 
of the $m^\text{\underline{th}}$ normal mode of the chain 
(it is not equal to $\hat A_m$), and 
\begin{equation}                                                                         \label{sesmc}
{\varsigma}_k = \sqrt{ \omega (\omega + \kappa) - 
                       \omega \kappa \cos\frac{(k - 1)  \,\pi}{ N} } 
\,\,\,\,\,\, 
(k = 1,...,N)
\end{equation}
are the eigenfrequencies of the normal modes.   
Also, the matrix ${\pmb O}$ arises from the diagonalization of the potential part 
of Hamiltonian (\ref{hamchain}), and it is given by \cite{nicacio4}
\begin{equation}                                                                         \label{matO}
{\pmb O}_{\! j k}  = 
\sqrt{\frac{2-\delta_{j1}}{N} } \, \cos \frac{(j-1)(2k-1) \,\pi}{2 N}.
\end{equation}
For the nonunitary part of the dynamics, 
the generator $\mathcal{\check L}(\check \rho)$ is found to be \cite{nicacio4} 
\begin{equation}                                                                         \label{lopthermal2}     
\begin{aligned} 
\mathcal{\check L}' (\check \rho) =& - \frac{1}{2}\zeta ({\bar n}_{\rm th} + 1)
\left( 
\{ {\hat{a}}_m^{'\dag} \hat{a}'_m , \hat \rho  \}
- 2 \hat{a}'_m \hat \rho {\hat{a}}_m^{'\dag}   \right)     \\
&- \frac{1}{2}\zeta\, {\bar n}_{\rm th}
\left( 
\{ \hat{a}'_m {\hat{a}}_m^{'\dag} , \hat \rho  \}
- 2 {\hat{a}}_m^{'\dag} \hat \rho \hat{a}'_m   \right), 
\end{aligned}
\end{equation}
where  
\begin{equation}
\hat{a}'_m := \sqrt{\frac{\varsigma_m}{\omega}} \frac{(\hat a_m + \hat a_m^\dagger)}{2} + 
              \sqrt{\frac{\omega}{\varsigma_m}} \frac{(\hat a_m - \hat a_m^\dagger)}{2}  
\end{equation}
is a squeezed annihilation operator related to $\hat a_m$. 

Finally, defining the collective vector
\begin{equation}                                                                         \label{vecx}
\hat x := (\hat{\sf q}_a, \hat {\sf q}_b, \hat {\sf q}_c,\hat Q_m, 
    \hat {\sf p}_a, \hat {\sf p}_b,\hat {\sf p}_c, \hat P_{m}), 
\end{equation}
we write the elements of the covariance matrix (CM) $\mathbf V$ of the system as 
\begin{equation}                                                                         \label{cmdef}
\mathbf V_{\! jk} (t)  =  
\frac{1}{2} {\rm Tr}\left[ 
                           \left\{ \hat x_j - \langle \hat x_j \rangle_t , 
                           \hat x_k - \langle \hat x_k \rangle_t \right\}_+
                           \check \rho(t)
                     \right],   
\end{equation}
where $\hat x_j $ is the $j^\text{\underline{th}}$ component of $\hat x$ 
and its mean value is
$\langle \hat x_j \rangle_t : = {\rm Tr}[ \hat x_j \check \rho(t) ]$. 
With the help of Eq.~(\ref{lindeff}), one can show that \cite{nicacio} 
\begin{equation}                                                                         \label{cmev}
\frac{d}{d t} \mathbf V = 
{ \bf \Gamma }\mathbf V + \mathbf V {\bf \Gamma}^\top + {\bf D} , 
\end{equation}
with  
\begin{equation}                                                                         \label{decmateff} 
\begin{aligned}                                                                
{\bf \Gamma } &:= - \frac{\zeta}{2} \mathsf I_{8} + 
\begin{pmatrix}
{\bf 0}_4 & \bf H \\
-\bf H    & {\bf 0}_4 &
\end{pmatrix}, \\                                                                           
{\bf D} & :=  \hbar \zeta({\bar n}_{\rm th} + \frac{1}{2}) 
\left( \mathsf I_3 \oplus \frac{\zeta_m}{\omega} \oplus 
       \mathsf I_3 \oplus \frac{\omega}{\zeta_m}          \right), 
\end{aligned}
\end{equation}
where $\mathsf I_{n}$ and ${\bf 0}_n$ are, respectively, 
the $n\times n$ identity and zero matrices, and
\begin{equation}                                                                         \label{hesseff}
\!\!\!\!{\mathbf H} :=\!
\left(\!\! \begin{array}{cccc}
\frac{\epsilon}{4} & 0 & 0 & 
-\frac{\epsilon \sqrt{\omega}}{4\sqrt{\varsigma_m}} \pmb{O}_{ \! m \alpha }\\
0 & \frac{\epsilon}{4} & 0 & 
-\frac{\epsilon \sqrt{\omega}}{4\sqrt{\varsigma_m}} \pmb{O}_{ \! m \beta } \\
0 & 0 & 0 & 0 \\
-\frac{\epsilon \sqrt{\omega}}{4\sqrt{\varsigma_m}} \pmb{O}_{ \! m \alpha } & 
-\frac{\epsilon \sqrt{\omega}}{4\sqrt{\varsigma_m}} \pmb{O}_{ \! m \beta } & 0 & 
\frac{\epsilon\omega  (\pmb{O}_{ \! m \alpha }^2 + \pmb{O}_{ \! m \beta }^2)  }
     { 4\varsigma_m }\\
 \end{array}\! \!\right)\! .
\end{equation}
%
By integrating Eq.~(\ref{cmev}), one finds
\begin{equation}                                                                         \label{cmsol1}
{\bf V}(t)  = {\rm e}^{-\zeta t} 
              {\mathsf E}_t \, {\bf V}\!_{0}  \, {\mathsf E}^\top_t + 
                    \frac{1}{\zeta} 
                    \left( 1 - {\rm e}^{-\zeta t}  \right) {\bf D}.  
\end{equation}
Note that 
$\mathsf E_t \in {\rm Sp}(8,\mathbb R) \cap {\rm O}(8)$, and its matrix
elements are shown in the \hyperlink{Appendix}{Appendix}. 
The simplicity of the effective dynamics given by Eq.~(\ref{cmsol1}) will enable us to 
obtain simple analytical results for the correlations. 
This is very important because with analytical expressions much of the physics becomes
apparent, something impractical or even impossible to do with the exact solution of the
original many-body problem. To gain critical insight into the validity of the 
simplified description, from which we draw physical conclusions, 
we will always present the numerical result obtained by solving the exact problem, {\it i.e.}, 
the propagation of the CM using the whole set of $N+3$ oscillators.  
%

\section{Quantum Correlations}\label{QC}
For pure bipartite quantum states, all correlations are associated with the ability to 
violate Bell inequalities, {\it i.e.}, a deviation from local realism, caused by the presence 
of entanglement \cite{gisin}. 
For mixed states, the situation is much more involved. 
For example, entanglement is required to violate Bell inequalities, 
but not all entangled states lead to such a violation \cite{bellmix}. 
In the same way, although pure-state entanglement guarantees quantum teleportation, 
the same does not occur with entangled mixed states \cite{belltele}. 
In order to deal with all these subtleties, the characterization of quantum correlations 
in mixed states usually employs the point of view of resource theory, 
in which correlations are seen, for instance, as resources for local 
transformation of states \cite{task}. 
In this paper, we want to understand how different correlations propagate through 
coupled harmonic oscillators. In addition to Bell-like nonlocalities and entanglement, 
already mentioned, we will also investigate mutual information, discord, and steering.

Consider the (column) vector 
$\hat {\sf x} := (\hat{q}_u,\hat{p}_u,\hat{q}_v,\hat{p}_v)^\dag$, 
composed by the position operators and the canonical conjugate momenta of 
two oscillators labeled $u$ and $v$. 
The Wigner function can be defined in terms of the mean value of 
the displaced parity operator \cite{grossmann,royer,ozorio},
$\hat R_{\xi} := \hat T_\xi \hat R_0 \hat T_\xi^\dagger$, 
evaluated at the phase-space point 
$\xi = (\xi^u_q,\xi^u_p,\xi^v_q,\xi^v_p ) \in \mathbb R^4$, 
\begin{equation}                                                                         \label{wig_0}
W(\xi) := 
\frac{1}{(\pi\hbar)^2} {\text {Tr} } \left[ \hat R_{\xi} \hat\rho_{[uv]} \right] , 
%
\end{equation}
where $\hat T_\xi^\dagger$ is the Heisenberg translation operator and 
$\hat R_0$ is the parity operator at the origin of the phase space. 
For Gaussian states, 
with {\it null} first moments of position and momentum, such as the ones used in 
this work, Eq.~(\ref{wig_0}) leads to
\begin{equation}                                                                         \label{wig}
W(\xi) = \frac{ \exp[-\frac{1}{\hbar} \xi^{\!\top} \,
              {\overline{\mathbf V}_{\![uv]}}^{\!\!-1} \xi  ]       } 
       { (\pi\hbar)^2 \sqrt{ \det \overline{\mathbf  V }_{\![uv]}}  },
\end{equation}
where
\begin{equation}                                                                         \label{cmbip}
{\overline{\mathbf V}}_{\![uv]} := 
\left(\begin{array}{cc}
 {\bf C}^{[u]}  & {\bf C}^{[uv]} \\
 {\bf C}^{[vu]} & {\bf C}^{[v]}
\end{array} \right)  
\end{equation}
is a $4 \times 4$ matrix in which ${\bf C}^{[u]} := {\bf C}^{[uu]} $ and 
\begin{equation}
\!\!{\bf C}^{[uv]}\!  := \! \frac{1}{\hbar} 
\left(\!\begin{array}{cc}
        {\rm Tr}[  \{ \hat q_u , \hat q_v \}_{\!_+} \hat \rho_{[uv]} ]  & 
        {\rm Tr}[  \{ \hat q_u , \hat p_v \}_{\!_+} \hat \rho_{[uv]} ]  \\
        {\rm Tr}[  \{ \hat p_u , \hat q_u \}_{\!_+} \hat \rho_{[uv]} ]  & 
        {\rm Tr}[  \{ \hat p_u , \hat p_v \}_{\!_+} \hat \rho_{[uv]} ]    
      \end{array}\! \right) . 
\end{equation}
Note the position-momentum reordering of the vector $\hat {\sf x}$ 
with respect to Eq.~(\ref{vecx}). 
Note also that the true CM of the system $[uv]$, like the one defined in (\ref{cmdef}), 
should be $\frac{\hbar}{2} {\overline{\mathbf V}}_{\![uv]}$. 
Again, we remember that the first moments are null for all times in the problem 
considered here.
Finally, before presenting the measures of correlations used in this work, 
in a suitable form for Gaussian states, it is convenient to present the local symplectic 
invariants $\mathcal{I}_{uv}$ and $\mathcal{I}_{u}$, defined as \cite{simon}  
\begin{equation}                                                                         \label{locinv0}
\mathcal{I}_{uv} := \det{\bf C}^{[uv]} = 
\mathcal{I}_{vu}, \,\,\, \mathcal{I}_{u} := \det{\bf C}^{[u]}. 
\end{equation}
%

The mutual information (MI) is widely used to quantify the 
total correlation (quantum and classical) 
in a state. This is because mutual information quantifies how far the 
joint probability for random variables deviate from the product of the 
marginals \cite{ollivier}. 
In this way, a state with null mutual information is not correlated in any way.  
For Gaussian bipartite states, the 
mutual information $M_{uv}$ is given by \cite{serafini}
\begin{equation}                                                                         \label{mi}
M_{uv} =  h( \sqrt{I_u} ) + h( \sqrt{I_v} ) 
                     -  h( \sigma^+_{\!uv} ) - h( \sigma^-_{\!uv} ),                        
\end{equation}
where                                                                         
\begin{equation}
h(x)  := \frac{1}{2} ( x + 1 ) \, {\rm ln}\frac{x + 1}{2} -                            
         \frac{1}{2} ( x - 1 ) \, {\rm ln}\frac{x - 1}{2},   
\end{equation}
and 
$\sigma^\pm_{\!uv}$ are the symplectic eigenvalues of 
${\overline{\mathbf V}}_{\![uv]}$ given by \cite{serafini}
\begin{equation}                                                                         \label{simpeig}
\sigma^\pm_{\!uv} = \sqrt{\frac{\Delta_{uv}}{2} \pm 
                    \frac{1}{2}
                    \sqrt{\Delta_{uv}^2 - 4 \det {\overline{\mathbf V}}_{\![uv]}}}, 
\end{equation}
with $\Delta_{uv} := \mathcal{I}_u + \mathcal{I}_v + 2 \mathcal{I}_{uv}$.  

From the difference between two classically equivalent definitions 
of the mutual information emerges the concept of quantum discord \cite{ollivier,henderson}. 
This correlation shows up with the existence of noncommuting observables 
in quantum mechanics, and it can be present even in nonentangled states.  
For pure states, both the mutual information and the discord 
are measures of entanglement {\it per se}. 
Operationally, discord can be seen as a figure of merit \cite{datta} for 
characterizing the resources needed for the computational protocol DQC1 \cite{knill}. 
Not only that, discord was proved also to be a prerequisite for the success of 
entanglement distribution through exchange of a carrier \cite{mauro_d}. 
In the bipartite Gaussian-state scenario, one can employ the so-called 
Gaussian quantum discord (GQD), which reads  \cite{giorda,pirandola}
\begin{equation}                                                                         \label{qdisc}
D_{\overrightarrow{uv}} = M_{uv} -  h( \sqrt{I_u} ) + 
               h\left(\!\!\frac{ \sqrt{\mathcal I_u} 
               + 2 \sqrt{\mathcal I_u \mathcal I_v } + 2 \mathcal I_{uv} } 
                        { 1 + 2\sqrt{\mathcal I_v} }\!\!\right),                        
\end{equation}
which measures the acquired information about the subsystem $u$ when 
measurements are performed on $v$. Note that, in general,
$D_{\overrightarrow{uv}} \ne D_{\overleftarrow{uv}} := D_{\overrightarrow{vu}}$.

The measure of entanglement considered here is the 
Logarithmic Negativity, which is based on the Peres-Horodecki criteria. 
For the bipartite Gaussian state in Eq.~(\ref{wig}), 
it reads
\begin{equation}                                                                         \label{ln}
E_{uv} = -
\min \left[ {\rm ln} \, \tilde\sigma^-_{\!uv}, 0 \right] ,
\end{equation}
where $\tilde\sigma^-_{\!uv}$ is the smallest symplectic eigenvalue of the matrix 
$\mathsf T_{\!\rm p} {\overline{\mathbf V}}_{\![uv]} \mathsf T_{\!\rm p}$, with 
$\mathsf T_{\!\rm p} := {\rm Diag}(1,1,1,-1)$. 
Alternatively, $\tilde\sigma^-_{\!uv}$ is obtained from (\ref{simpeig}) 
replacing $\mathcal I_{uv} \to - \mathcal I_{uv}$, {\it i.e.}, 
replacing $\Delta_{uv}$ by 
$\tilde{\Delta}_{uv} := \mathcal I_{u} +\mathcal I_{v} - 2\mathcal I_{uv}$.

The ability to change the reduced state of part of a composed system by acting on 
another part of that same system is a kind of correlation usually called 
Quantum Steering (QS) \cite{wiseman}. 
If, by allowing local measurements and classical communication, one part of the system 
is able to convince the other part that they share an entangled state, one says 
the state is steerable. 
Operationally, this power of persuasion is translated into a way of providing  
security in a one-sided quantum key distribution protocol \cite{branciard}. 
Unlike entanglement, but very much like GQD, 
steering is not a symmetric correlation. In other word, two parts of the 
same system always share the same amount of entanglement while, 
in general, they share different amounts of steering, 
which is quantified as \cite{kogias}
\begin{equation}                                                                         \label{steer}
S_{\overrightarrow{uv}} =   \frac{1}{2} \max\left[{\rm ln} \, \mathcal I_{u} 
- {\rm ln}\,{\det{\overline{\mathbf V}}_{\![uv]}} , 0 \right]
\end{equation}
for the two-mode Gaussian state in (\ref{wig}). 
Changing the roles of $j$ and $k$, 
the extension $ S_{\overleftarrow{jk}} := S_{\overrightarrow{kj}} $ measures 
the steering in the other direction. 

Finally, we want to investigate correlations which allow quantum mechanics 
to violate local realism. 
Given the dichotomic nature of the displaced parity operator \cite{grossmann,ozorio}, 
which has the spectrum $\{+1,-1\}$, it turns out to be possible to construct 
a Bell-CSCH inequality based on measurements of its eigenvalues \cite{bellparity}. 
By writing  $\hat R_{\xi} = \hat R_{\xi^u}\hat R_{\xi^v}$, where 
$\xi^u := (\xi^u_q,\xi^u_p)$ and $\xi^v := (\xi^v_q,\xi^v_p)$, 
we define the ``Bell operator'' \cite{bellparity}
\begin{equation}\label{oi}
{\hat B} := \hat R_{\xi^u}\hat R_{\xi^v}  + \hat R_{\xi^u}\hat R_{\xi'^v} +
             \hat R_{\xi'^u}\hat R_{\xi^v} - \hat R_{\xi'^u}\hat R_{\xi'^v}.  
\end{equation}
A violation of local realism occurs when the state $\hat \rho$ 
of a bipartite system is such that $|{\rm Tr}({\hat B}  \hat \rho) |> 2$. 
Using (\ref{wig}) and the locality of the parity operator, 
$W(\xi) = W(\xi^u,\xi^v) = {\rm Tr}( \hat R_{\xi^u}\hat R_{\xi^v}  
                                     \hat \rho_{[uv]})/(\pi\hbar)^{2} $ 
is the two mode Wigner function of a bipartite Gaussian state. 
Explicitly, 
\begin{equation}                                                                         \label{bell1}
\begin{aligned}
\langle \hat B \rangle& = 
(\pi\hbar)^{2}  W(\xi^u,\xi^v) +          (\pi\hbar)^{2} W(\xi^u,\xi'^v) \\  
& + (\pi\hbar)^{2}  W(\xi'^u,\xi^v) -  (\pi\hbar)^{2}  W(\xi'^u,\xi'^v). 
\end{aligned}
\end{equation}
This expression is a function of eight real parameters, which makes maximization of $B$ a
hard problem, in general. What is usually done, is to limit the problem to a subset 
of the phase space. 
A possible choice, used in this work, is $\xi^u = \xi^v = (0,0)$ and 
$\xi'^u = - \xi'^v = (\theta,0)$. The problem is then reduced to the  
search of $\theta$ that maximizes $\langle \hat B \rangle$, {\it i.e.}, 
we investigate Bell nonlocality through
\begin{equation}                                                                         \label{bell2}
 \bar{B}_{uv} = 
\underset{\{\theta \in \mathbb R\}}{\max}{|\langle \hat B \rangle|}, 
\end{equation}
which signalizes a legitimate quantum correlation as soon as it exceeds a value of 2. 
With this, we finish our ``bestiary'' of correlation measures. 

Before going to the results, 
it is worth remarking that the set of steerable states is a subset of the entangled states 
and that the set of nonlocal Bell states is a subset of the steerable states \cite{jones}. 

\section{ Transport of Entanglement and Steering }\label{TQC}     
%
We begin our study by considering all oscillators, except $c$ and $b$, 
in the product state of local vacuum states. In this way, there are no correlations 
at all between these oscillators. On the other hand, we will be considering that, 
initially, the oscillators $c$ and $b$ are prepared in a two-mode thermal 
squeezed state (TMTSS) \cite{marian}. 
The CM in Eq.~(\ref{cmdef}) for this initial global state reads 
\begin{equation}                                                                         \label{cm0}
{\bf V}_{\!0} = \frac{\hbar}{2} 
\left( \begin{array}{cccc}
       1 & 0 & 0 & 0\\
       0 & s & w & 0\\
       0 & w & y & 0\\
       0 & 0 & 0 & 1
       \end{array}  \right)  \oplus 
\frac{\hbar}{2} 
\left( \begin{array}{cccc}
       1 & 0 & 0 & 0\\
       0 & s & -w & 0\\
       0 & -w & y & 0\\
       0 & 0 & 0 & 1
       \end{array}  \right)        
\end{equation}
with
\begin{equation}
\begin{aligned}
s &:= 2 \bar n_c \sinh^2 \frac{r}{2} + \cosh r , \\
y &:= 2 \bar n_c \cosh^2 \frac{r}{2} + \cosh r , \\
w &:= (\bar n_c + 1) \sinh r, 
\end{aligned} 
\end{equation}
where $r \ge 0$ is the squeezing parameter and $\bar n_c \ge 0$ is the mean number of 
thermal excitations in oscillator $c$. By simply collecting the elements corresponding 
to the subsystems of interest ($[ab]$, $[ac]$, etc.), 
one can use  ${\bf V}_{\!0}$ to construct $\overline{{\bf V}}_{\![uv]}$ 
as defined in (\ref{cmbip}). 

In this section, we will present our results concerning propagation of steering and 
entanglement. In order to characterize these properties in the initial state of the 
system, we use  Eqs.~(\ref{locinv0}) and~(\ref{cm0})  to obtain the invariants
\begin{eqnarray}                                                                         \label{locinv}
&&\mathcal{I}_b = [ (1 + \bar n_c)\cosh r - \bar n_c]^2,  \nonumber \\ 
&&\mathcal{I}_{bc} = - (\bar n_c + 1)^2 \sinh^2 r, \nonumber \\ 
&&\mathcal{I}_c = [ (1 + \bar n_c)\cosh r + \bar n_c]^2.
\end{eqnarray}
In addition to the fact that $\det\overline{{\bf V}}_{\![bc]} = (2 \bar n_c + 1)^2$, 
the use of these invariants to evaluate Eqs.~(\ref{ln}) and~(\ref{steer}) leads to 
\begin{eqnarray}                                                                         \label{incorr}
E_{bc}^{t=0} &=& \max \left[ 0, - {\rm ln}\,{\tilde \sigma^-_{\!bc}}  
                         \right],                                                        \nonumber \\
S_{\overrightarrow{\overleftarrow{bc}}}^{t=0} &=& \max\left[ 0,   
{\rm ln} \left|\frac{\bar{n}_c \mp (1 + \bar{n}_c) \cosh r}{1 + 2 \bar{n}_c}\right|
\right],     
\end{eqnarray}
respectively, where 
\begin{eqnarray}                                                                         \label{spl1}
&&\!\!\!\!\!\!\!\!\!({\tilde \sigma^-_{\!bc}})^2 =  
2 (\bar n_c + 1)^2 \cosh^2 r - (2 \bar n_c + 1)                                          \nonumber \\                    
&&                   - 2(\bar{n}_c + 1)\cosh r \sqrt{ (\bar{n}_c + 1)^2 
                                                       \cosh^2 r - (2\bar{n}_c + 1) }.
\end{eqnarray}%
The initial state of $\hat \rho_{bc}$ is entangled for $r > 0$ and 
any value of ${\bar n}_c$, since ${\tilde \sigma^-_{\!bc}} <1$ \cite{marian}.
  
%
\subsection{Unitary Evolution}
%
The temporal evolution governed by Eq.~(\ref{lindblad}) will mix, 
distribute, and rearrange the initial correlations of the pair $[bc]$. 
Considering the ideal unitary case, {\it i.e.}, 
in the absence of reservoirs, the effective model of the evolution of an 
initial state with CM ${\bf V}_{\!0}$, governed by Eq.~(\ref{cmsol1}) with $\zeta = 0$, 
becomes  
\begin{equation}                                                                         \label{cmsol2}
{\bf V}(t)  =  \mathsf E_t \, {\bf V}\!_{0}  \, \mathsf E_t^\top.
\end{equation}
Using the matrix elements of $\mathsf E_t$ shown in the \hyperlink{Appendix}{Appendix}, 
one obtains for the pair $[ac]$  
\begin{eqnarray}                                                                         \label{funct}
\det \overline{{\bf V}}_{\![ac]} &=& 
\left[  2 \bar{n}_c + 1 - (\bar{n}_c + 1) ( \cosh r -1) ( 2 F -1  ) \right]^2 \!,        \nonumber \\
\mathcal{I}_a &=& 
\left[ 1 + 2 (\bar{n}_c + 1 ) ( \cosh r-1) F  \right]^2,                                 \nonumber \\
\mathcal{I}_c &=& [\bar{n}_c + (\bar{n}_c + 1) \cosh r]^2,                               \nonumber \\
\mathcal{I}_{ac} &=& -2  (\bar{n}_c + 1 )^2 F \sinh^2 r ,                                            
\end{eqnarray}
where $F = F(t)$ is defined in (\ref{auxfunc}) and it only depends 
on the properties of the chain through (\ref{sesmc}) and (\ref{matO}).
Analogously, for the pair $[bc]$, one finds  
\begin{eqnarray}                                                                         \label{funct2}
\det \overline{{\bf V}}_{\![bc]} &=& 
\left\{ (2\bar{n}_c + 1) G + 2[\bar{n}_c + (\bar{n}_c +1)\cosh r] H \right.              \nonumber \\
&& \left. - \,  2 ( \bar{n}_c + 1 )  (\cosh r-1)I\right\}^2,                             \nonumber \\
\mathcal I_b &=& 
\left\{ [\bar{n}_c - (\bar{n}_c +1)\cosh r] G - 2H \right.                               \nonumber \\
&& \left. - 2 (\bar{n}_c + 1 ) ( \cosh r-1) I \right\}^2,                                \nonumber \\
\mathcal{I}_{bc} &=& - (\bar{n}_c + 1)^2 (G + 2I) \sinh^2 r ,                                      
\end{eqnarray}
where $G,H,I$ are defined in (\ref{auxfunc}) to depend only on 
the chain properties through (\ref{sesmc}) and (\ref{matO}). 
From these three equations, only $I=I(t)$ is time dependent.
With all these in hand, we can perform an analytical investigation of 
the correlations defined in Eqs.~(\ref{ln}) and~(\ref{steer}).

We start by having a look at the critical points of the entanglement 
and steering as functions of time. 
From Eq.~(\ref{simpeig}), it is possible to show that a critical 
point $t_\star$ of $\tilde\sigma^-_{\!uv}$ obeys 
\begin{equation}                                                                         \label{crit}
\left.\left[
\frac{d}{dt} \det{\overline{{\bf V}}_{\![uv]} } + 
\left(\tilde\sigma^-_{\!uv}\right)^2 
\frac{d}{dt}\tilde\Delta_{uv}\right] \right|_{t = t_\star} = 0;  
\end{equation}
{\it i.e.}, the simultaneous critical point of $\det{\overline{{\bf V}}_{\![uv]} }$ and 
$\tilde\Delta_{uv}$ gives the solution for $d \tilde\sigma^-_{\!uv}/dt = 0$. 
%
%
%
Moreover, it is easy to check that the critical points of 
$\det \overline{{\bf V}}_{\![ac]}$ and  
$\Delta_{ac}$ coincide with those of the function $F$ in (\ref{auxfunc}), 
and that they are solutions of the transcendental equation
\begin{equation}                                                                         \label{cpt}
\sin\left(\chi \tau \right) = \sin \tau + \sin[(\chi-1) \tau] 
\end{equation}
for 
$\chi = \frac{\omega}{\varsigma_m}{\pmb O}_{m \alpha}^2 + 
\frac{\omega}{\varsigma_m}{\pmb O}_{m \beta}^2 + 1$, and  
$\tau = \frac{\epsilon t}{4}$. 
The critical points do not depend on the 
initial state (\ref{cm0}) in any way. As 
 a matter of fact, they depend only on the structure of the chain 
which is contained in the diagonalizing matrix $\pmb O$.    

The dependence on the initial state enters when we evaluate the correlations at the 
specific critical times. For instance, the entanglement measure in Eq.~(\ref{ln}), 
when evaluated at a critical point, reads
\begin{equation}                                                                         \label{entcrit}
E_{ac}^{t = t_\star} = 
\frac{1}{2} {\rm ln}\frac{(1+\bar{n}_c)\sinh^2 r}{\cosh r - 1} 
             - \frac{1}{4} {\rm ln}
              \left.\left[\det\overline{{\bf V}}_{\![ac]} \right]\right|_{t = t_\star}.             
\end{equation}
A similar analysis can be done to the pair $[bc]$, 
but it should take into account the time dependence of the function $I$, 
which has the fixed points determined by solutions of
\begin{equation}                                                                         \label{cpt2}
\left( \frac{ {\pmb O}_{m \beta} }{ {\pmb O}_{m \alpha} }\right)^2 
\sin\left(\chi \tau \right) =  -\sin \tau - \sin[(\chi-1) \tau] . 
\end{equation}
For ``direct'' $S_{\overrightarrow{uv}}$ and ``reverse'' $S_{\overleftarrow{uv}}$ 
in (\ref{steer}), it is not difficult to demonstrate that the critical points are 
once again given by Eqs.~(\ref{cpt}) and~(\ref{cpt2}), which are, respectively, 
the critical points of $\mathcal I_a$ and $\mathcal I_b$. 
From these, we can conclude that the positions of the maxima are the {\it same} in 
the entanglement and steering dynamics. This will be clear 
in the plots shown later in this paper.

\subsubsection{Initial Pure States}\label{ueps}
In order to make sure that the treatment using the effective description 
(\ref{lindeff}),  from which analytical treatment is possible, indeed produces trustful 
results, we will also present numerical solutions obtained by solving the 
global dynamics (\ref{lindblad}). 
The latter will be referred to as the exact solution, but one should 
understand that this is so only in the sense of a numerical approach. 
The problem (\ref{lindblad}) is not, in general, amenable to exact analytical treatment.   

\begin{figure}[!b]                                                                                                                        
\includegraphics[width=8cm,trim = 0 20 0 0]{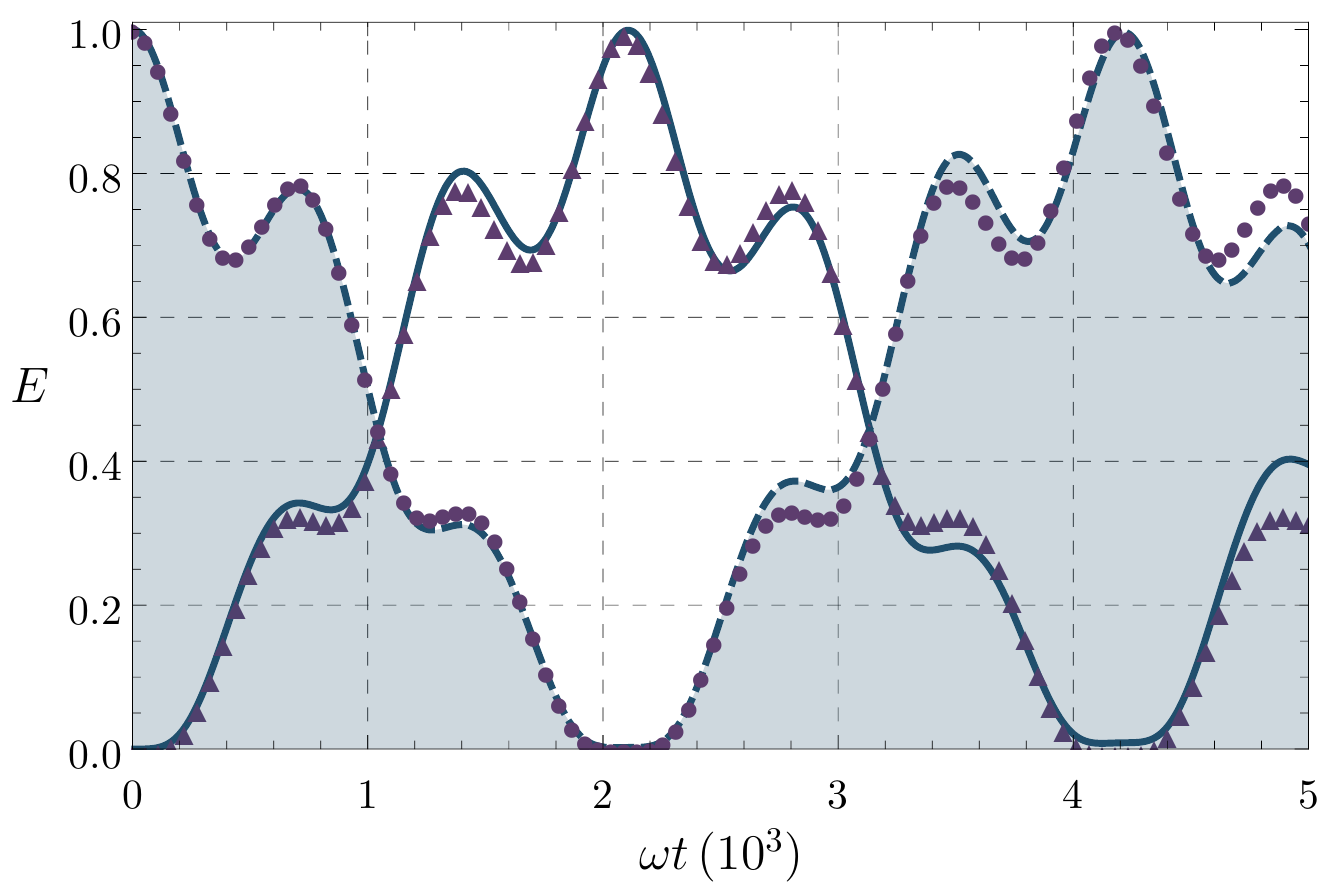} 
\caption{(Color Online) %
Unitary dynamics of entanglement as a function of dimensionless time $\omega t$. 
The dashed curve plus shading and the solid line correspond to the exact time evolution 
of $E_{bc}$ and $E_{ac}$, respectively. 
Circles and triangles are, respectively, $E_{bc}$ and $E_{ac}$ evaluated within the 
effective model.
The chain consists of $N = 10$ oscillators, and the chosen system parameters are 
$\kappa/\omega = 20$, $\alpha = N$, $\beta = 1$, 
$\epsilon/\omega = 0.03$, $\Omega/\omega = 1$. With this choice, there is resonance 
with the mode $m = 1$, {\it i.e.}, $\varsigma_1 = \Omega = \omega$. 
For the initial state, we choose $r = 1$ and $\bar{n}_c = 0$.
%
}                                                                                        \label{fig2ent1}                                                      
\end{figure}

\begin{figure}[!htbp]                                                                                                                        
\includegraphics[width=8cm,trim = 0 20 0 0]{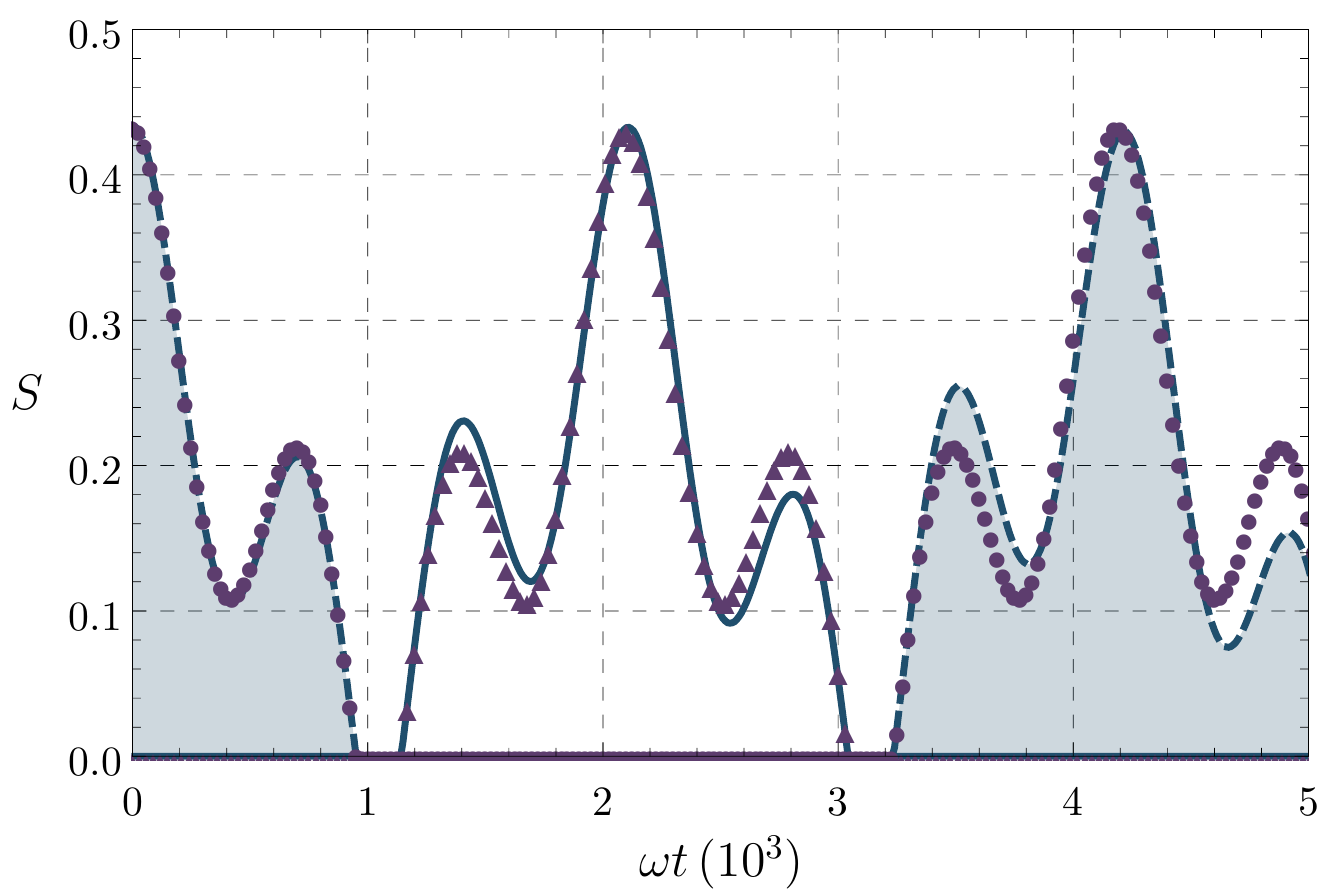}   
\caption{(Color Online) %
Unitary dynamics of ``direct'' steering as a function of 
dimensionless time $\omega t$. 
The parameters are the same as those in Fig.~\ref{fig2ent1}. 
The dashed curve plus shading and the solid line correspond to 
the exact time evolution of $S_{\protect\overrightarrow{bc}}$ 
and $S_{\protect\overrightarrow{ac}}$, respectively. 
Circles and triangles are, respectively, 
$S_{\protect\overrightarrow{bc}}$ and $S_{\protect\overrightarrow{ac}}$ 
evaluated within the effective model.
}                                                                                        \label{fig3steer1}                                                      
\end{figure}

\begin{figure}[!htbp]                                                                                                                        
\includegraphics[width=8cm,trim = 0 20 0 0]{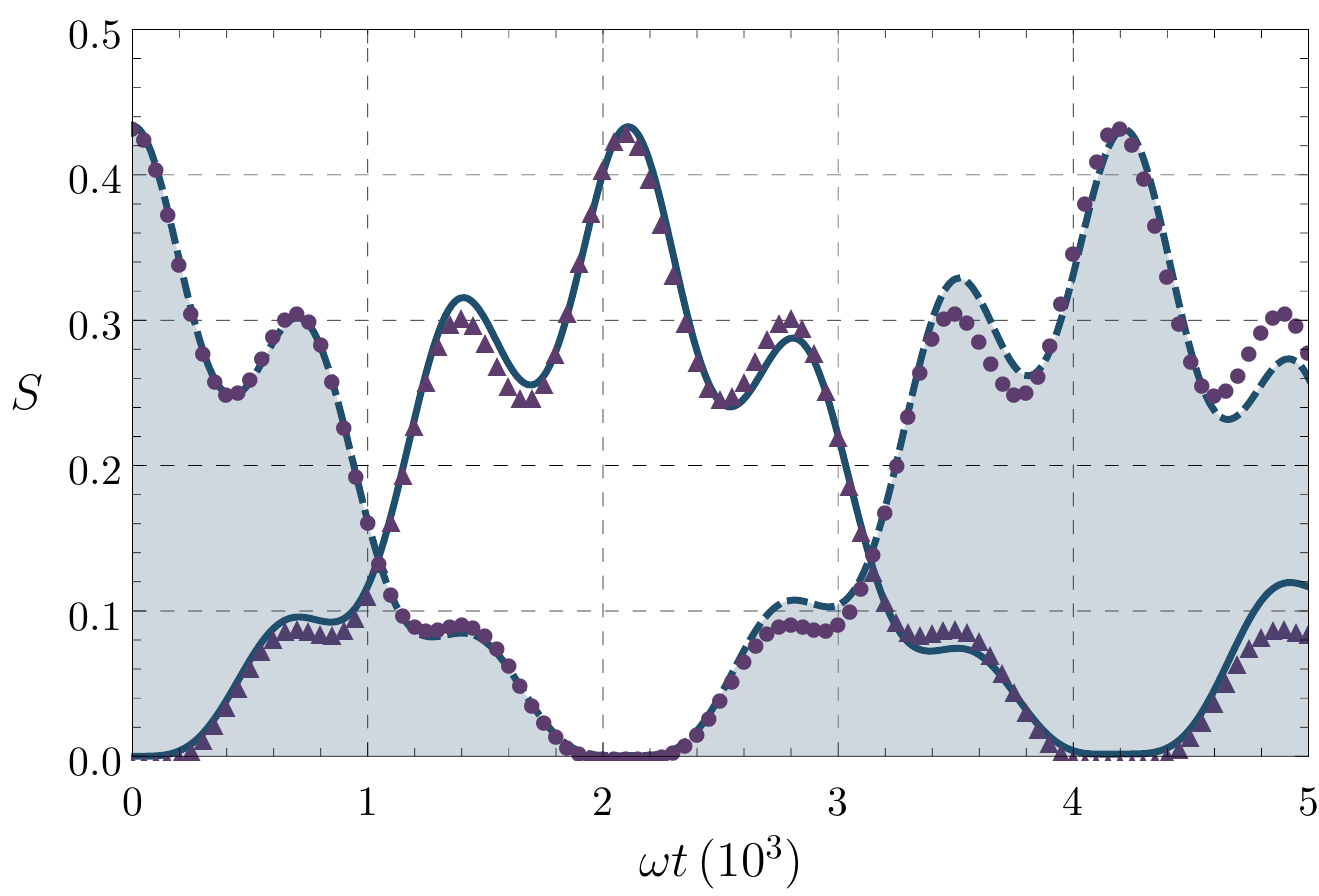}   
\caption{(Color Online) %
Unitary dynamics of  ``reverse'' steering as a function of 
dimensionless time $\omega t$. 
The parameters are the same as those in Fig.~\ref{fig2ent1}. 
The dashed plus shading and the solid line correspond to 
the exact time evolution of $S_{\protect\overleftarrow{bc}}$ 
and $S_{\protect\overleftarrow{ac}}$, respectively. 
Circles and triangles are, respectively, 
$S_{\protect\overleftarrow{bc}}$ and $S_{\protect\overleftarrow{ac}}$ 
evaluated within the effective model.   
}                                                                                        \label{figsteer2}                                                      
\end{figure}

For the unitary case, the results for the effective 
model come from the use of Eqs.~(\ref{funct}) and~(\ref{funct2}). 
We start our study by taking $\bar{n}_c = 0$, such that the global initial 
state (\ref{cm0}) is pure and, according to Eq.~(\ref{incorr}), 
one finds $E_{bc}^{t=0}=r$. All other pairs of oscillators are initially not entangled.
As time passes, the initial entanglement in $[bc]$ is progressively transferred to the 
pair $[ac]$ and back \cite{plenio2005}, 
as seen from Fig.~\ref{fig2ent1}. 
``Direct''  and ``reverse'' steerings are shown in 
Figs.~\ref{fig3steer1} and \ref{figsteer2}, respectively.   
It is worth noticing that zero entanglement implies zero steering in any direction, 
while zero steering does not necessarily imply zero entanglement. 
This is because the set of steerable states is a subset of the set 
of entangled states \cite{jones}.
%
%

%
For the parameters used to produce the plots in Fig.~\ref{fig2ent1}, 
the initial entanglement in $[bc]$ is practically fully transferred to the pair $[ac]$. 
Such highly efficient transfer is observed in Figs.~\ref{fig3steer1} and \ref{figsteer2} 
for steering as well. However, an interesting situation now arises.
Although the initial ``direct'' and ``reverse'' steerings are equal
when $\bar{n}_c = 0$, see Eq.~(\ref{incorr}), their dynamics are interestingly different. 
In our simulations, the `reverse'' steering dynamics is closer to entanglement dynamics 
than the ``direct''  counterpart. One can see, for instance, that $S_{\overleftarrow{ac}}$ 
behaves much more like $E_{ac}$  than  $S_{\overrightarrow{ac}}$. 
The same is true for the pair $[bc]$.   
Furthermore, the ``direct'' steering $S_{\overrightarrow{bc}}$ suffers
sudden death and sudden rebirth (see Fig.~\ref{fig3steer1}),  
while neither the ``reverse'' steering $S_{\overleftarrow{bc}}$ nor entanglement does so, 
see Figs.~\ref{fig2ent1} and \ref{figsteer2}.

Finally, it is interesting to notice that the transfer of ``direct'' steering shown in 
Fig.~\ref{fig3steer1} is also abrupt or {\it sudden}. 
The correlation $S_{\overrightarrow{bc}}$ is allowed to arise only after the sudden death 
of $S_{\overrightarrow{bc}}$. 
Not only that, there are finite time intervals in which both steerings are null. 
%
\subsubsection{Initial Mixed States}\label{uems}
\begin{figure}[!b]                                                                                                                        
\includegraphics[width=8cm,trim = 0 20 0 0]{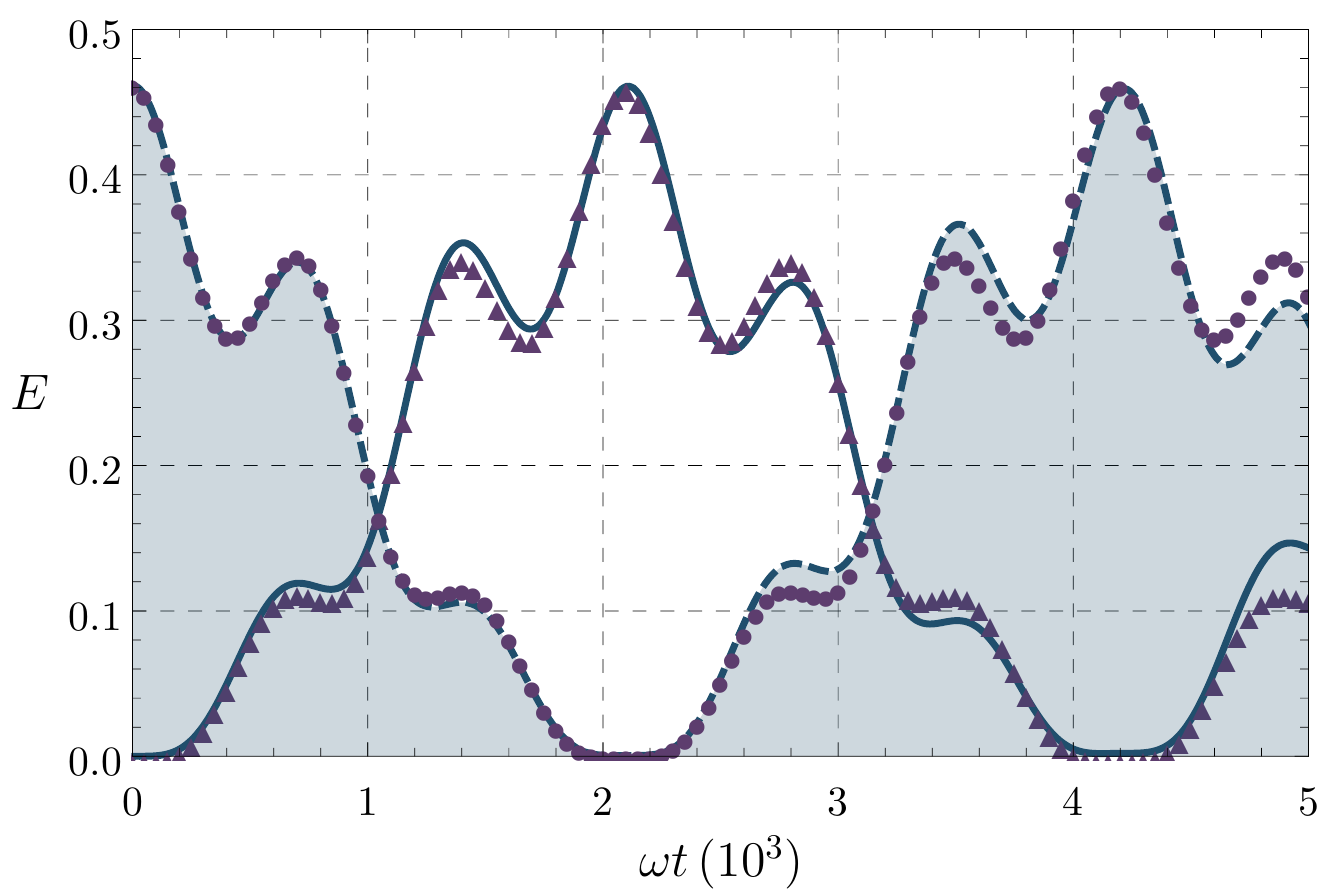}                                                   
\caption{(Color Online) %
Unitary dynamics of entanglement as a function of dimensionless 
time $\omega t$. 
The dashed curve plus shading and the solid line correspond to the 
exact time evolution of the $E_{bc}$ and $E_{ac}$, respectively. 
Circles and triangles are, respectively, $E_{bc}$ and $E_{ac}$ 
evaluated within the effective model.
For the initial state, we chose $r = 1$ and $\bar{n}_c = 10$, 
and the other parameters are the same as in Fig.~\ref{fig2ent1}.           
}                                                                                        \label{fig5ent2}                                                      
\end{figure}

According to Eq.~(\ref{incorr}), we know that an increase in the thermal 
occupation number $\bar n_c$ in oscillator $c$ must degrade the total entanglement 
initially present in the system. This is clearly seen from Fig.~\ref{fig5ent2} by 
comparing it to Fig.~\ref{fig2ent1}.  In general, the entanglement dynamics stays 
qualitatively similar to the unitary case, including the high efficiency. 
This has been numerically investigated for a broad range of $r$ and $\bar n_c$. 
We found that the efficiency of the transport is practically perfect for 
$  0 \le r \le 5 $ and $ 0 \le \bar n_c \le 50$, 
at the critical point $\omega t_\star = 2094.4$ 
[calculated using Eq.~(\ref{cpt}) 
with $N = 10$, $\alpha = 1$, $\beta = 10$, and $ m = 1 $].  
As mentioned, the critical point is also the maximum of the function $F$ 
defined in Eq.~(\ref{auxfunc}).   
These numerical results suggest that the maximum value in (\ref{entcrit}) 
is exactly equal to $E_{bc}^{t=0}$, but we have not managed to prove it. 
%

\begin{figure}[!htbp]                                                                                                                        
\includegraphics[width=8cm,trim = 0 20 0 0]{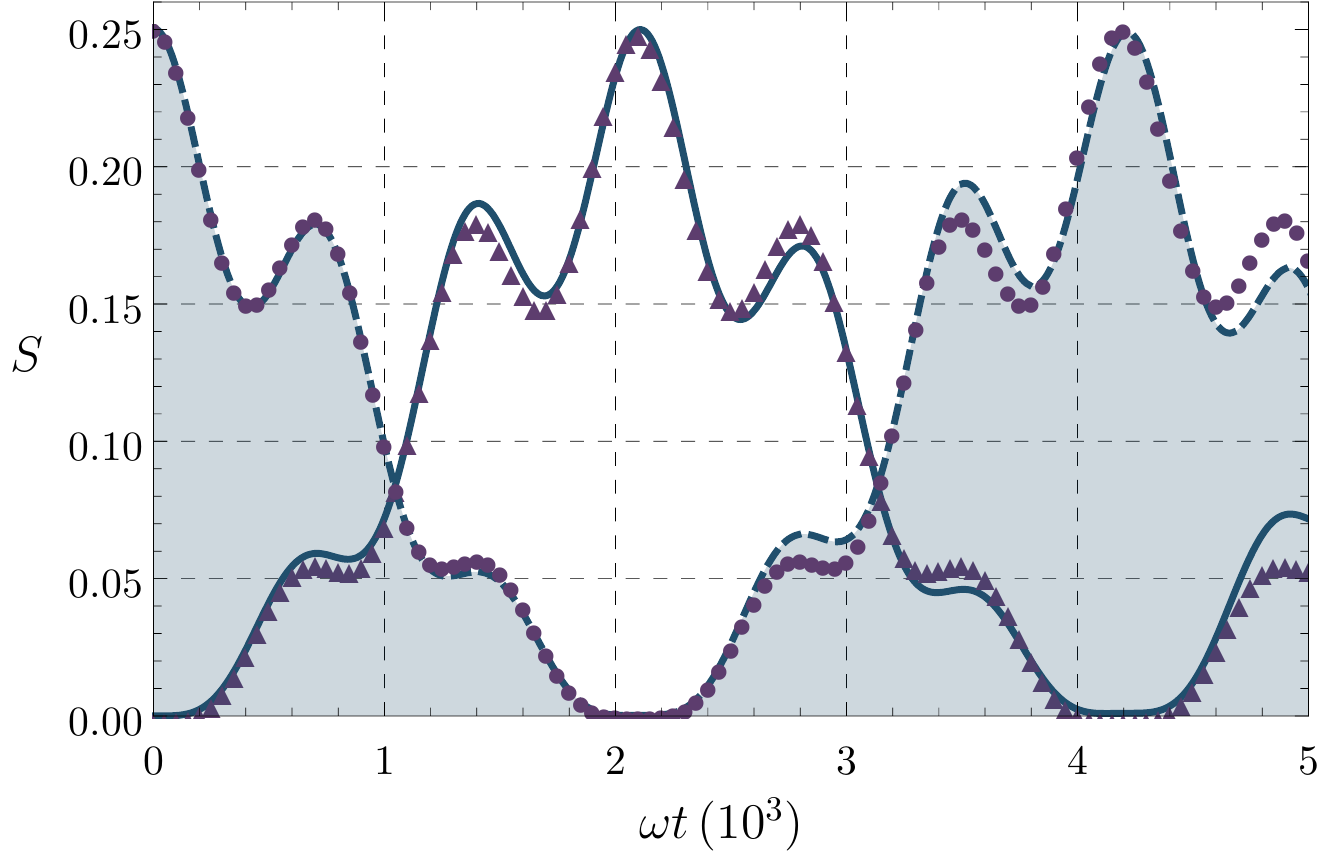}                                                   
\caption{(Color Online) Unitary dynamics of  ``reverse'' steering as 
a function of dimensionless time $\omega t$. 
The physical situation is the same as in Fig.~\ref{fig5ent2}. 
The dashed curve plus shading and the solid line correspond to the exact 
time evolution of $S_{\protect\overleftarrow{bc}}$ and 
$S_{\protect\overleftarrow{ac}}$, respectively. 
Circles and triangles are, respectively, 
$S_{\protect\overleftarrow{bc}}$ and $S_{\protect\overleftarrow{ac}}$ 
evaluated within the effective model.
}                                                                                         \label{fig6steer3}                                                      
\end{figure}

On the other hand, the steering presents a richer dynamics due to the possibility of 
non
equivalence between the ``direct'' and ``reverse'' cases. This is already manifested in 
the initial state since, according to Eq.~(\ref{incorr}), 
$ S_{\overrightarrow{bc}}^{t=0} = 0 $ for 
$ r \le r_{\rm c} = {\arccosh}[ (3 \bar{ n }_c + 1)/(\bar{ n }_c + 1) ] $,  while 
$S_{\overleftarrow{bc}}^{t=0}$ is non null. It is remarkable that the ``direct'' 
steering is not created dynamically, {\it i.e.}, 
$S_{\overrightarrow{ac}}^t = 0\,\,\, \forall t $, as we could numerically conclude. 
In other words, there was no ``transmutation between species'' of steering.
In Fig.~\ref{fig6steer3}, we show that the ``reverse'' initial steering, 
$S_{\overleftarrow{bc}}^{t=0} \ne 0 $, is totally transferred to the pair $[ac]$. 
It is important to report that for values of $r$ greater than the critical 
value $r_{\rm c}$, the ``reverse'' steering does not undergo deaths and rebirths 
similar to Fig.~\ref{fig6steer3},  while the ``direct'' steering (not shown) does. 
%

From (\ref{steer}) and (\ref{funct}), 
it is easy to show that the ``direct'' steering $S_{\overrightarrow{ac}}$ 
is zero provided 
\begin{equation}                                                                         \label{zeroac}
F \le \frac{1}{4} + \frac{{\bar n}_c}{2({\bar n}_c + 1)(\cosh r -1)}.  
\end{equation}
The equality is observed when there is sudden death or rebirth of 
$S_{\overrightarrow{ac}}$.  In spite of the fact that the LHS of (\ref{zeroac}) 
is a highly transcendental function, it does not depend on the details of the initial 
state, {\it i.e.}, $r$ and $\bar n_c$, and it can, in principle, be numerically solved. 
For the parameters used in Fig.~\ref{fig3steer1}, the use of Eq.~(\ref{zeroac}) 
allows one to predict that $S_{\overrightarrow{ac}}$ will be non-null in the 
interval $ 1143.1\le \omega t  \le 3045.8$.  
The condition for the nullity of the ``direct steering'' $S_{\overrightarrow{bc}}$ 
can be obtained in a similar way, but it is much more involved since it involves 
the functions $G$, $H$, and $I$. 
However, it can be handled, and we applied it to find that $S_{\overrightarrow{bc}}$ will 
vanish in the interval $951.3\le \omega t  \le 3237.5$. 
This is in complete agreement with the plots shown in Fig.~\ref{fig3steer1}.  

The next natural step is to include thermal excitations in the other 
oscillators as well. For this, let us consider the CM (\ref{cmdef}) 
\begin{equation}                                                                         \label{cm02}
\!\!{\bf V}'_{\!0} \!=\! \hbar (\bar n + \tfrac{1}{2}) \! \left[\!
\left(\! \begin{array}{cccc}
       1 & 0 & 0 & 0\\
       0 & s & w & 0\\
       0 & w & s & 0\\
       0 & 0 & 0 & 1
       \end{array} \! \right) \!  \oplus \!
\left( \! \begin{array}{cccc}
       1 & 0 & 0 & 0\\
       0 & s & -w & 0\\
       0 & -w & s & 0\\
       0 & 0 & 0 & 1
       \end{array} \!\! \right)  \right],    
\end{equation}
where $s := \cosh r$, $w := \sinh r$, 
$r \ge 0$ is the squeezing parameter, and 
$\bar n \ge 0$ is the mean number of thermal photons of each oscillator. 
The initial entanglement and steering now read
\begin{eqnarray}                                                                         \label{incorr2}
E_{bc}^{t=0} &=& \max[ 0, r - {\rm ln}(2 \bar n + 1)],  \nonumber \\
S_{\overrightarrow{bc}}^{t=0} &=&  S_{\overleftarrow{bc}}^{t=0}  = 
\max\left[ 0, {\rm ln} \left( \frac{\cosh r}{ 2 \bar{n} +1}  \right) \right],  
\end{eqnarray}
from which it follows that the state of $b$ and $c$ will be entangled only when \cite{marian}
\begin{equation}                                                                         \label{sepcond}
r >  {\rm ln} [(2\bar n + 1)].      
\end{equation}
By the same token, that state will be steerable only when
$ r > \arccosh( 2 \bar n + 1 )$. Note that 
${\bf V}'_{\!0} = (2\bar{n} + 1) {\bf V}_{\!0}|_{\bar{n}_c = 0}$, 
with ${\bf V}_{\!0}$ in Eq.~(\ref{cm0}), such that the application of Eq.~(\ref{ln}) results in
\begin{equation}                                                                         \label{ln2}
\!\!\!\!E_{uv}({\bf V}'_{\!0}) = 
\max \left[ 0, E_{uv}({\bf V}_{\!0}|_{\bar{n}_c = 0}) 
            - {\rm ln}(2\bar{n} + 1) \right]. 
\end{equation}
From Eq.~(\ref{steer}),
\begin{equation}                                                                         \label{steer2}
S_{\overrightarrow{\overleftarrow{uv}}}({\bf V}'_{\!0}) = \max\left[0 , 
S_{\overrightarrow{\overleftarrow{uv}}}({\bf V}_{\!0}|_{\bar{n}_c = 0}) 
- {\rm ln}(2\bar{n} + 1) \right]. 
\end{equation}

The immediate conclusion is that the presence of the thermal excitations reduces 
uniformly both the initial entanglement and steering 
by a factor of ${\rm ln}(2\bar{n} + 1)$ when compared to the pure state 
${\bf V}_{\!0}|_{\bar{n}_c = 0}$.  
The evolution of these two quantities is presented in 
Figs.~\ref{fig2ent1}, \ref{fig3steer1}, and \ref{figsteer2}. 
From these, a finite $\bar n$ indeed affects the total correlations available 
to propagate, but the transfer process itself remains very efficient. 
The explanation relies on the fact that all time dependence 
is still contained only in the functions $F$ and $I$, which do not depend on 
$r$ or $\bar n$. 
Given that there are no qualitative differences from the previous simulations, 
we do not show plots for this case.
%
\subsection{Interaction with the Environment}
To take into account the influence of the environment on the transport of correlations, 
the dynamics of the CM will now be described by Eq.~(\ref{cmsol1}) in the effective model 
and by Eqs.~(\ref{lindblad}) and (\ref{lopthermal}) in the exact model.
The analysis of Eq.~(\ref{cmsol1})  reveals that the influence of the initial state, 
through  ${\mathbf V}_0$, is progressively attenuated with characteristic time 
$\zeta^{-1}$. 
In other words, the reservoirs progressively erase the initial correlations in the system.
This can be seen from the asymptotic state,  
\begin{equation}                                                                         \label{sstate2}
\!\!\!\lim_{t\to \infty} \check{\bf V}(t)  = \frac{1}{\zeta}\check{\bf D} = 
\hbar (\bar n_{\rm th} + \tfrac{1}{2}) \!
\left(\! \mathsf I_3 \oplus \frac{\zeta_m}{\omega} \oplus 
       \mathsf I_3 \oplus \frac{\omega}{\zeta_m}          \!\right),
\end{equation}
which is completely uncorrelated.  

\begin{figure}[!b]                                                                                                                        
\includegraphics[width=8cm,trim = 0 20 0 0]{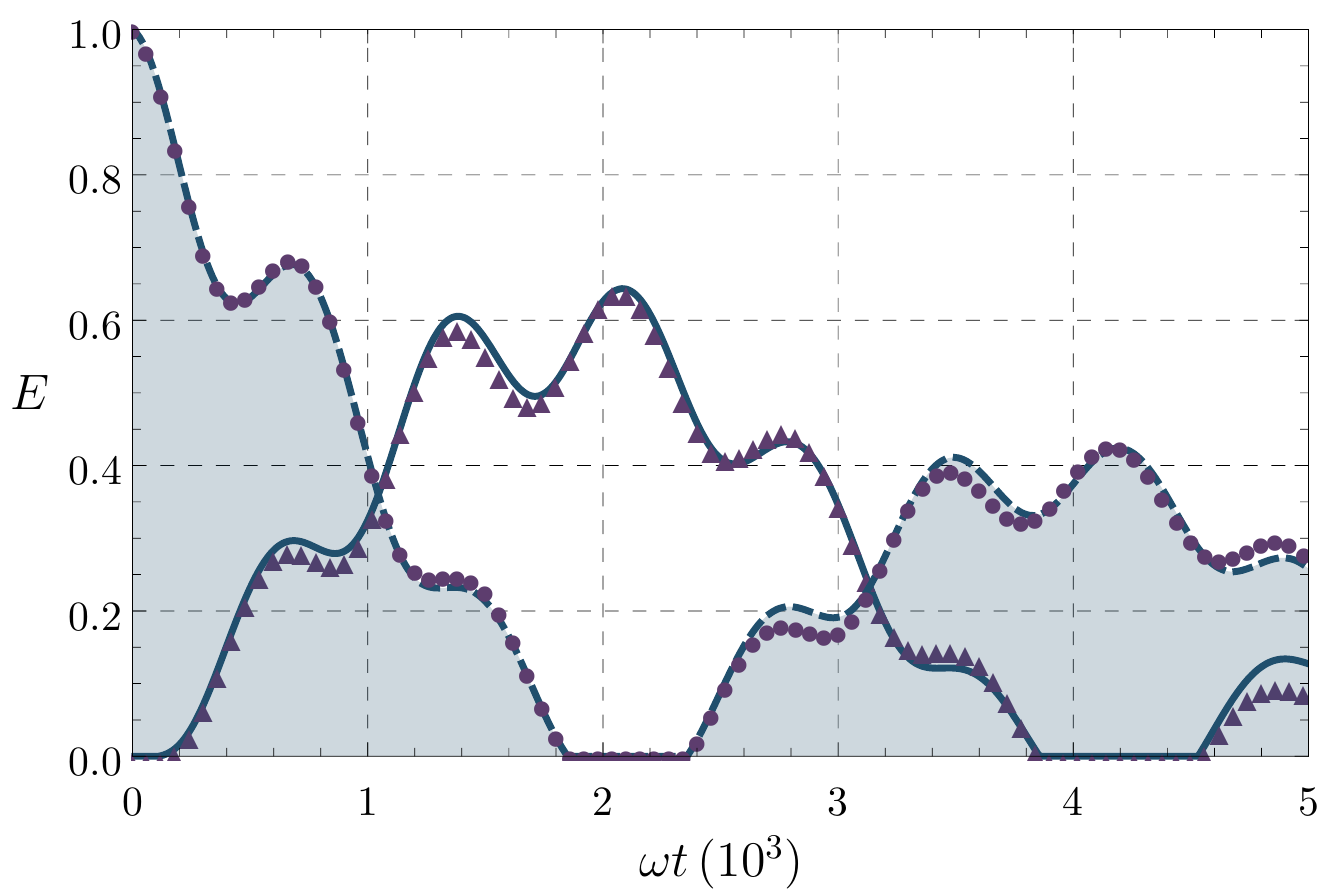}    
\caption{(Color Online) %
Open system dynamics of entanglement as a function of dimensionless 
time $\omega t$. 
The dashed curve plus shading and the solid line correspond to the 
exact time evolution of $E_{bc}$ and $E_{ac}$, respectively. 
Circles and triangles are, respectively, $E_{bc}$ and $E_{ac}$ 
evaluated within the effective model.
The parameters are the same as in Fig.~\ref{fig2ent1}, 
except for the new ones that concern interaction with the 
reservoirs. They are chosen here to be $\bar{n}_{\rm th} = 0.1$ 
and $\zeta/\omega = 10^{-4}$.                 
}                                                                                        \label{fig7ent3}                                                      
\end{figure}

\begin{figure}[!t]         
\includegraphics[width=8cm,trim = 0 20 0 0]{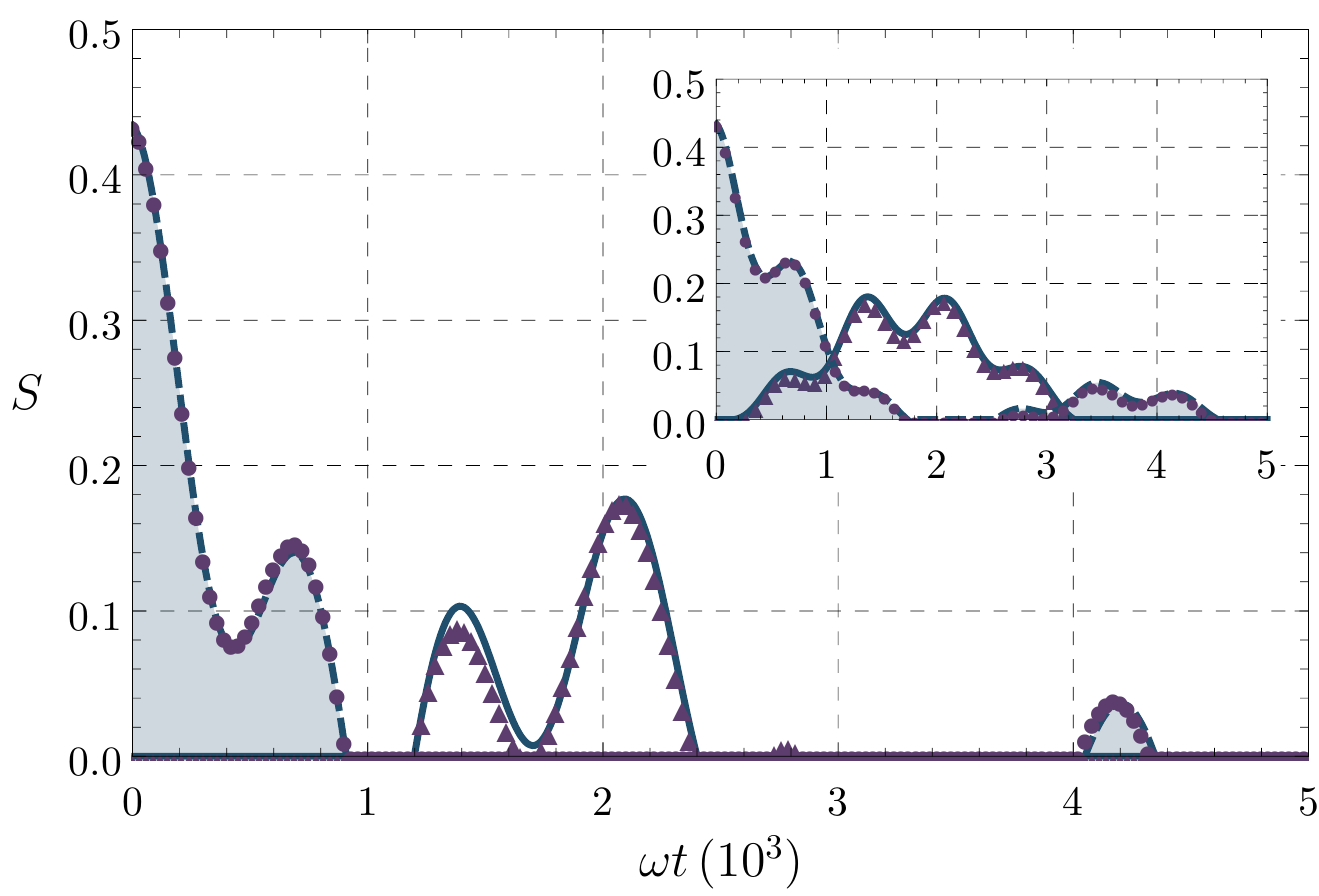} 
\caption{(Color Online) Open system dynamics of ``direct'' steering 
as a function of dimensionless time $\omega t$. The parameters are 
the same as those in Fig.~\ref{fig7ent3}. 
The dashed curve plus shading and the solid line correspond to the exact 
time evolution of $S_{\protect\overleftarrow{bc}}$ and 
$S_{\protect\overleftarrow{ac}}$, respectively. 
Circles and triangles are, respectively, 
$S_{\protect\overleftarrow{bc}}$ and $S_{\protect\overleftarrow{ac}}$ 
evaluated within the effective model. 
Inset: The same for ``reverse'' steering.    
The dashed curve plus shading and the solid line correspond to 
the exact time evolution of $S_{\protect\overrightarrow{bc}}$ and 
$S_{\protect\overrightarrow{ac}}$, respectively. 
Circles and triangles are, respectively, 
$S_{\protect\overrightarrow{bc}}$ and 
$S_{\protect\overrightarrow{ac}}$ evaluated within the effective model. 
}                                                                                        \label{fig8steer4}                                                      
\end{figure}

In Figs.~\ref{fig7ent3} and {\ref{fig8steer4}}, we plot the dynamical behavior of  
logarithmic negativity and steering, respectively. 
It is clear that both are progressively destroyed, as predicted by Eq.~(\ref{sstate2}). 

%
\section{Mutual Information and Quantum Discord } \label{miaqd}       
We now analyze the propagation of mutual information and quantum discord initially 
present in the pair $[bc]$. Let us start once again by considering an initial state 
in which the pair $[bc]$ is found in a TMTSS, while the remaining oscillators are 
prepared in a product of local vacuum states. In the sequence, we will consider 
thermal excitations in all oscillators, and we will also discuss the effect of thermal 
reservoirs. 

\begin{figure}[!htbp]                                                                                                                        
\includegraphics[width=8cm,trim =0 20 0 0]{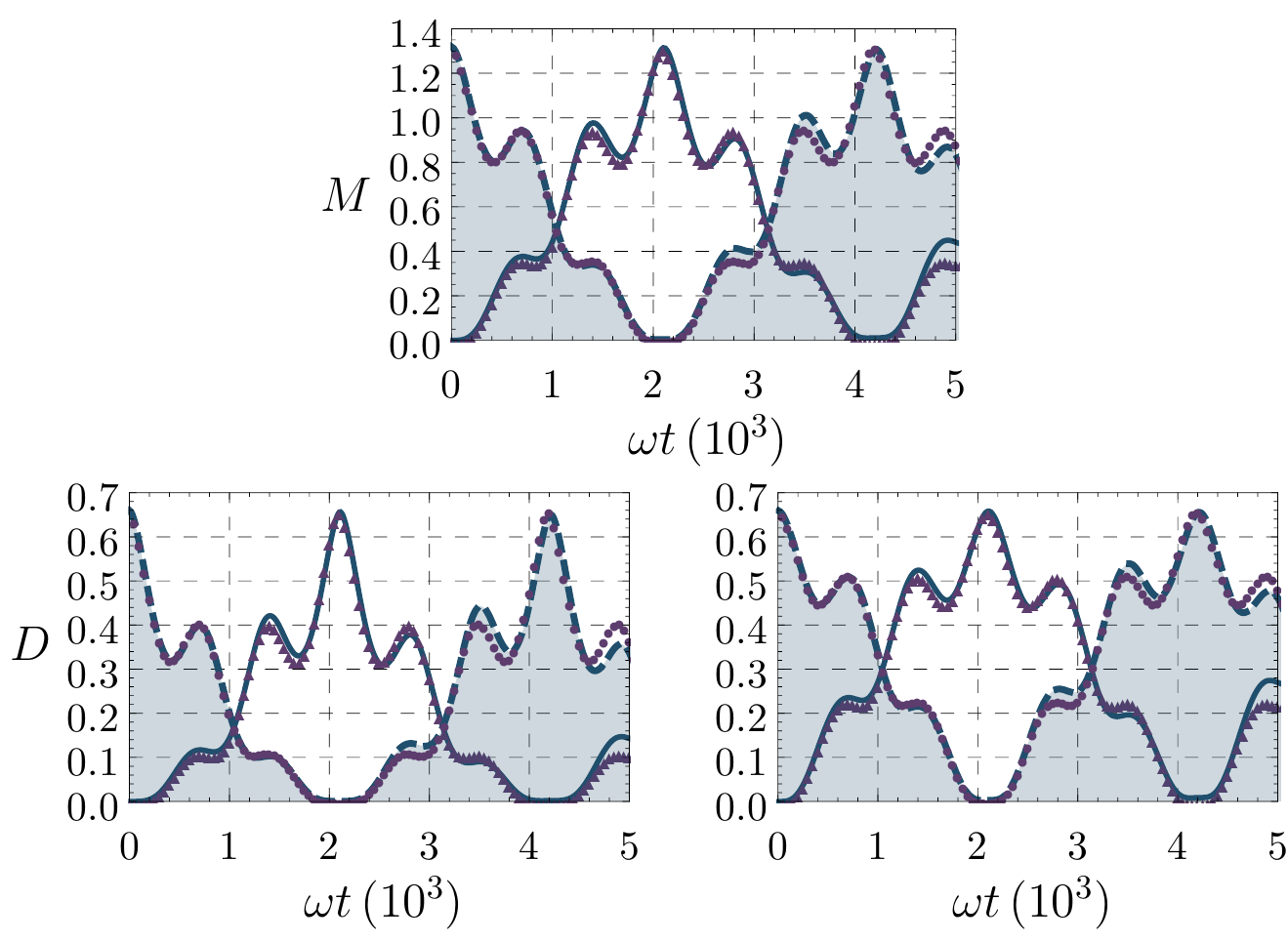}                                                   
\caption{(Color Online) %
Unitary dynamics of  Mutual Information and Quantum Discord as a function of 
dimensionless time $\omega t$. 
The parameters are the same as those of Fig.~\ref{fig2ent1}. 
{\bf Top:} 
The dashed curve plus shading and the solid line correspond to 
the exact time evolution of $M_{bc}$ 
and $M_{ac}$, respectively. 
{\bf Bottom Left:} 
The dashed curve plus shading and the solid line correspond to 
the exact time evolution of ``direct'' discords  $D_{\protect\overrightarrow{bc}}$ 
and $D_{\protect\overrightarrow{ac}}$, respectively.
{\bf Bottom Right:} 
The dashed curve plus shading and the solid line correspond to 
the exact time evolution of ``reverse'' discords  $D_{\protect\overleftarrow{bc}}$ 
and $D_{\protect\overleftarrow{ac}}$, respectively. 
For all plots, circles and triangles are 
the respective effective models.                             
}                                                                                        \label{fig9midisc1}                                                      
\end{figure}

For the unitary evolution, and CM given by Eq.~(\ref{cm0}), we can use the local 
invariants in Eq.~(\ref{locinv}) to calculate the initial amount of mutual information 
and discord, using Eqs.~(\ref{mi}) and (\ref{qdisc}), respectively. 
These will be propagated to the pair $[ac]$. The dynamics of these correlations 
is presented in Fig.~\ref{fig9midisc1}.  
It is interesting to see that, just like steering, these correlations have the 
same critical points and the efficiency of propagation as entanglement, 
see Fig.~\ref{fig2ent1}. 
Also, in spite of the fact that discord too is asymmetric with respect to the parties, 
there is no qualitative difference in their behavior like for steering 
(compare Figs.~\ref{fig3steer1} and \ref{figsteer2}). 
Indeed, both kinds of discord behave much like the entanglement with 
no sudden death or rebirth.  
Note also that when the mutual information vanishes, 
the other correlations do as well, while the converse is not true. 
This corroborates the fact that mutual information contains both classical 
and quantum correlations.  

\begin{figure}[!htbp]                                                                                                                        
\includegraphics[width=8cm,trim =0 20 0 0]{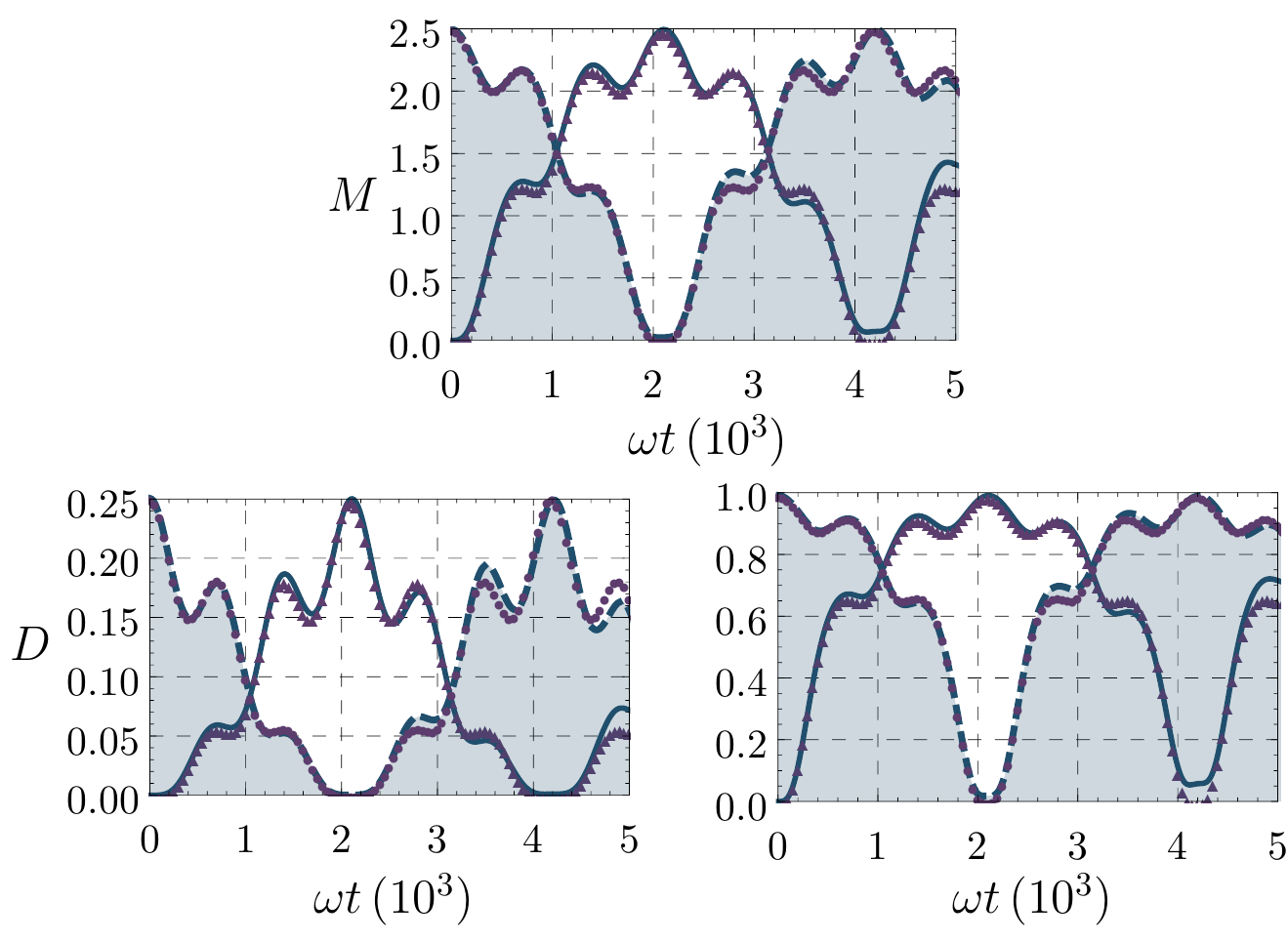}                                                   
\caption{(Color Online)  
Unitary dynamics of  Mutual Information and Quantum Discord as a 
function of dimensionless time $\omega t$. 
The physical situation is the same as in Fig.~\ref{fig5ent2}.  
The description of the curves is as in Fig.~\ref{fig9midisc1}.
}                                                                                        \label{fig10midisc2}                                                      
\end{figure}

By heating oscillator $c$, the initial mutual information, unlike entanglement,  
increases with $\bar n_c$. This is a result of the dependence on $\bar n_c$ which 
is found in Eq.~(\ref{locinv}). Interesting enough, while ``direct'' discord diminished 
with the increase of $\bar n_c$, the ``reverse'' discord did the opposite, 
compare Figs.~\ref{fig9midisc1} and Fig.~\ref{fig10midisc2}. 
Consequently, some quantumness is benefited by the increase of $\bar n_c$. 
One can also notice from  Fig.~\ref{fig10midisc2} that the efficiency of transport 
was not affected by thermal excitations initially present in oscillator $c$.


\begin{figure}[!t]                                                                                                                                                                                                                                            
\includegraphics[width=8cm,trim =0 20 0 0]{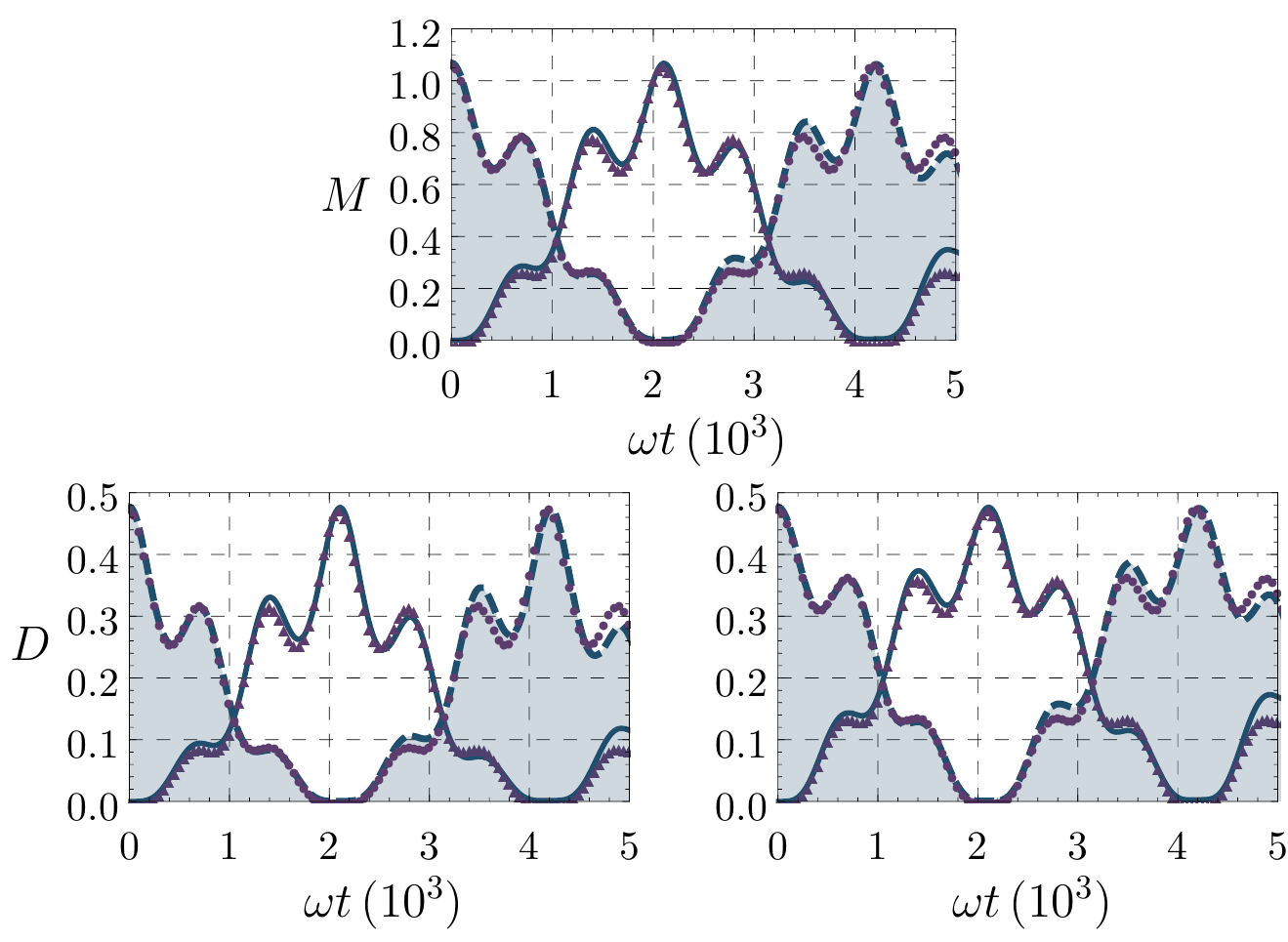}                                                   
\caption{(Color Online) 
Unitary dynamics of  Mutual Information and Quantum Discord as a function of 
dimensionless time $\omega t$.                                                     
The description of the curves is as in Fig.~\ref{fig9midisc1}, 
and the frequencies and coupling constants are the same as in Fig.~\ref{fig2ent1}.
All oscillators start in local thermal states with mean occupation number 
$\bar n = 10$, 
except for the pair $[bc]$, 
which is in a TMTSS with $r=1$ and the same number of 
thermal excitations as the other oscillators.   
The initial state for the effective model has the CM in 
Eq.~(\ref{cm02}) with $r = 1$ and $\bar{n} = 10$. 
}                                                                                        \label{fig11midisc3}                                                      
\end{figure}

Now we consider thermal excitations in all oscillators, just like at the end 
of Sec.~\ref{uems}. 
The effective model uses the initial CM defined in Eq.~(\ref{cm02}), and we found
\begin{eqnarray}                                                                         \label{locinv2}
\mathcal{I}_b &=& \mathcal{I}_c =  (2 \bar n+1)^2\cosh^2 r, \nonumber \\ 
\mathcal{I}_{bc} &=& - (2 \bar n+1)^2 \sinh^2 r,
\end{eqnarray}
which are used to calculate the initial mutual information and discord in the pair $[bc]$. 
The evolution of these correlations is plotted in Fig.~\ref{fig11midisc3}, 
where one can see that the available mutual information and discord (both directions)
decreased with $\bar n$ (compare to Fig.~\ref{fig9midisc1}).
It is interesting to notice that the initial state is symmetric with respect to discord; 
that is, at $t=0$ we have $D_{\protect\overrightarrow{bc}}=D_{\protect\overleftarrow{bc}}$. 
This happens because, for the initial state characterized by Eq.~(\ref{cm02}), 
one finds $\mathcal I_b = \mathcal I_c$. 
However, as time passes, this symmetry is dynamically lost. 
For the regime considered here and in spite of the presence of thermal excitations,
the high efficiency of the propagation of mutual information and discord is preserved, 
just like what happened to entanglement and steering. 

If we let the chain  interact with environmental thermal baths, the time evolution 
of the CM is given by Eq.~(\ref{cmsol1}). 
Basically, our simulations (not shown) indicated that mutual information and discord 
behaved very much like the entanglement, as already shown in Fig.~\ref{fig7ent3}. 
To be more specific, they dynamically decrease with a time scale of $\zeta^{-1}$ 
until completely vanishing in the asymptotic limit, see Eq.~(\ref{sstate2}). 
For this reason, we chose not to show plots for this case.

\begin{figure}[!htbp]                                                                                                                        
\includegraphics[width=8cm, trim= 0 20 0 0]{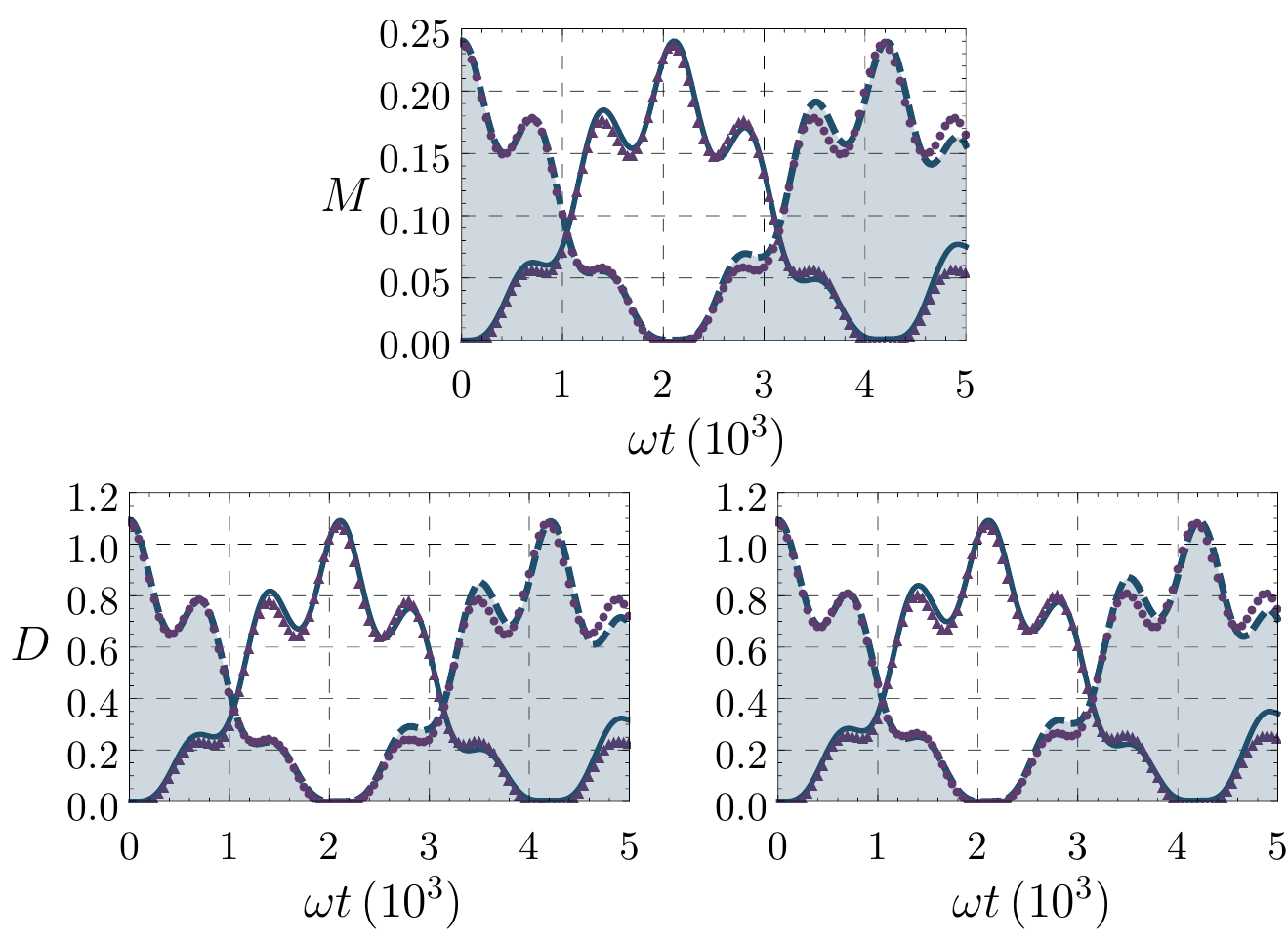}                                                   
\caption{(Color Online) %
Unitary dynamics of  Mutual Information and Quantum Discord as a function of 
dimensionless time $\omega t$.  The physical situation is the same as in 
Fig.~\ref{fig11midisc3}, except for 
the initial state of the pair $[bc]$, which is now 
separable. The initial CM used in the effective model is given by   
Eq.~(\ref{cm02}) with $r = 1/2$ and $\bar{n} = 10$.                                                                                         
}                                                                                        \label{fig12midisc2}                                                      
\end{figure}

In all situations analyzed so far, the propagation of correlations 
took place in a scenario where entanglement was always present. 
However, it is possible to consider initial states in which there is no entanglement 
but there are other forms of quantum or classical correlations present. 
We are interested in exploring this problem now, {\it i.e.}, we want to investigate 
propagation of these other-than-entanglement correlations in a situation with no entanglement. 
For this aim, we consider the initial state with the CM given by Eq.~(\ref{cm02}), 
with the choice $r < {\rm ln} [(2\bar n + 1)] $, see Eq.~(\ref{sepcond}). 
As one can see in Fig.~\ref{fig12midisc2}, 
transport is still equally efficient, even in the absence of entanglement.

\newpage

\section{Propagation of Bell Correlations} \label{pobc}               
We now focus on the propagation of Bell-like correlations. 
As previously discussed, they are related to the idea of violation of local realism.
To be more specific, we employ Eq.~(\ref{bell2}) as a quantifier of this type of 
correlation. 
At this point, one must have in mind that no violation of one form of the Bell 
inequality does not necessarily exclude violation of local realism in other configurations, 
{\it i.e.}, other inequalities built with results from different measurements. 
For the purposes of this paper, it is enough to investigate propagation of only one 
type of Bell-like correlation, which is the one quantified by Eq.~(\ref{bell2}). 

\begin{figure}[!htbp]                                                                                                                        
\includegraphics[width=8.5cm,trim = 0 20 0 0]{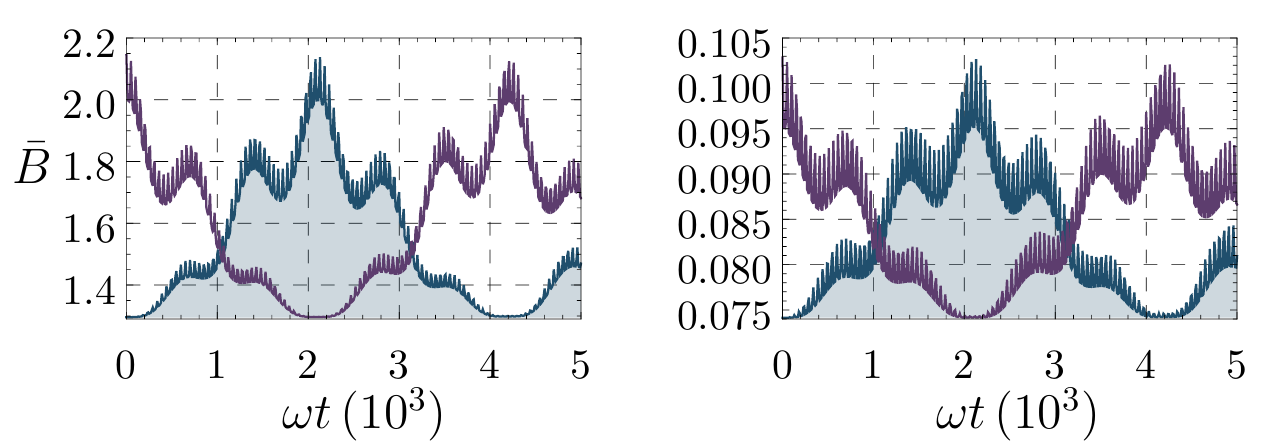}                                                   
\caption{(Color Online) %
Unitary dynamics of Bell correlations as a function of 
dimensionless time $\omega t$. 
The curve without shading is the exact time evolution of $\bar B_{ac}$, 
while the curve with the shading is the exact time evolution 
of the $\bar B_{bc}$. 
We avoided the comparison with the effective model 
due to the presence of very small oscillations, 
which makes visualization problematic.  
{\bf Left:} Unitary pure-state evolution as in Fig.~\ref{fig2ent1}.   
{\bf Right:} Unitary mixed-state evolution as in Fig.~\ref{fig5ent2}.     
}                                                                                        \label{fig13bell1}                                                      
\end{figure}

We consider the case of unitary evolution (already discussed) and two initial preparations. 
In one case, the state is pure, and in the other there are thermal excitations 
in oscillator $c$ (mixed state). 
These preparations have also been previously considered, {\it e.g.}, 
in Figs.~\ref{fig2ent1} and \ref{fig5ent2}, respectively. 
The dynamics is shown in  Fig.~\ref{fig13bell1}. 
Note that violation for the pair $[ac]$ is attained around $\omega t = 2 \times 10^3$ 
for the pure-state case (left panel). 
Just like before, this kind of correlation  is also efficiently transmitted through the chain. 
It is worthwhile to notice that, during the violation, the state is both entangled and steerable, 
see Figs.~\ref{fig2ent1}, \ref{fig3steer1} and \ref{figsteer2}. 
This is expected from the well-known hierarchy discussed earlier and presented in \cite{jones}.
By increasing the thermal energy in oscillator $c$, it comes to a point at which the correlations 
are not enough to violate local realism. This is shown in the right panel of  Fig.~\ref{fig13bell1}. 
It is interesting to see that the setup still works efficiently to propagate the correlations. %

The interaction with thermal baths does not bring much new physics for the propagation of this 
kind of correlation. As before, it is progressively destroyed on a time scale $\zeta^{-1}$. 
This destructive effect is expected, giving the fact that  $\bar B_{uv}$ depends on Wigner functions, 
whose deleterious impact, as consequence of the action of thermal baths, is well known. 
%
 
\section*{Final Remarks}\label{conc}                                
We provided a detailed study of the problem of propagation of correlations through a system 
of coupled harmonic oscillators. We considered propagation in the closed system scenario 
as well as in the presence of heat reservoirs. In addition, we studied the effect of having 
initially pure or mixed entangled states or even non entangled mixed states with non-null 
discord. The presence of thermal excitations in the system was also investigated.
We chose a particular configuration to this study which is well known to be reducible to 
a three-body system. This allowed us to provide analytical results that were useful for 
understanding the rich behavior found in the propagation of different correlations. 
We could, for instance, spot fundamental differences in the dynamical behavior of 
steering and discord, which are quantities that can be defined in non equivalent ways 
in the same bipartite system.  

It is important to remark that we performed simulations for a variety of chain lengths $N$ 
(smaller or bigger than the one considered here) and the same qualitative behavior was found. 
Of course, $N$ cannot be deliberately increased since this brings the chain spectrum to a 
continuum what spoils the transfer considerably. This was previously shown in the context of 
entanglement in \cite{plenio2005}. 
To the best of our knowledge, this is the first work to bring together entanglement, 
steering, mutual information, discord, and Bell-type correlations to a comparative study 
from the important point of view of quantum transport.
We hope this work can motivate further investigation in other setups, 
such as the ones found in coupled quantum electrodynamic systems, 
in which two-level systems can couple with harmonic oscillators, forming polarons. 
Another possibility is to use more intricate topologies of coupled systems using 
the general simplification approach presented in \cite{nicacio4}. It would be interesting to 
study the propagation of discord and steering in these different scenarios to see whether or not 
the asymmetry, present in the definition of these correlations, can also be pinpointed in those 
physical systems.

\acknowledgments 
FN and FLS are supported by the CNPq ``Ci\^{e}ncia sem Fronteiras''
program through the ``Pesquisador Visitante Especial'' initiative
(Grant No. 401265/2012-9).
FLS is a member of the Brazilian National Institute of Science and Technology of
Quantum Information (INCT-IQ) and acknowledges partial support from CNPq
(Grant No. 307774/2014-7).
%

\hypertarget{Appendix}{\section*{Appendix}} \appendix                 
\renewcommand{\theequation}{A-\arabic{equation}}  \setcounter{equation}{0}
The symplectic matrix $\mathsf E_t$ in (\ref{cmsol1}) is written as 
\begin{equation}
{\mathsf E}_t = 
\left(
\begin{array}{rl}
 \mathbf C  & \mathbf S \\
-\mathbf S  & \mathbf C
\end{array} \right),
\end{equation}
with the matrices $\mathbf C := \cos({\bf H} t) $ and 
                  $\mathbf S  := \sin({\bf H} t) $ for ${\bf H}$ in Eq.~(\ref{hesseff}).
Explicitly, their elements are                  
\begin{eqnarray}
{\mathbf C}_{11} &=&  \frac{\omega}{\varsigma_m}
\frac{   \pmb{O}_{ \! m \alpha }^2 [( \chi -1 ) \, + \,  \cos( \chi \tau ) ] \, + \,  
          \chi \, \pmb{O}_{ \! m \beta }^2 \cos \tau }
      { ( \chi - 1 ) \chi },                                                             \nonumber \\
{\mathbf C}_{12} &=& 
\frac{\omega}{\varsigma_m}\frac{   \pmb{O}_{ \! m \alpha }\pmb{O}_{ \! m \beta} 
           [  (\chi -1 ) \,  - \,  \chi \cos\tau \, + \,  \cos(\chi\tau) ]   }
      {  ( \chi - 1 ) \chi} ,                                                            \nonumber \\
{\mathbf C}_{21} &=& {\mathbf C}_{12}, \nonumber\\
{\mathbf C}_{14} &=& 2\sqrt{\frac{\omega}{\varsigma_m}}\frac{{\pmb O}_{m\alpha}}{\chi} 
                     \sin^2\left(\frac{ \chi \tau }{2} \right),                         \nonumber \\
{\mathbf C}_{22} &=&
\frac{\omega}{\varsigma_m}\frac{   \chi\, \pmb{O}_{ \! m \alpha }^2 \cos\tau \,  + \,  
               \pmb{O}_{ \! m \beta}^2 [ (\chi-1) \, + \, \cos(\chi \tau) ] }
      { ( \chi - 1 )\chi} ,                                                              \nonumber \\
{\mathbf C}_{24} &=& {\mathbf C}_{42} = 
2 \sqrt{\frac{\omega}{\varsigma_m}}\frac{ \pmb{O}_{ \! m \beta}}{\chi} 
\sin^2\left(\frac{\chi \tau }{2} \right) ,                                              \nonumber \\
{\mathbf C}_{13} &=& {\mathbf C}_{31} = {\mathbf C}_{23} = {\mathbf C}_{32} = 
{\mathbf C}_{34} = {\mathbf C}_{43} = 0,                                                 \nonumber \\
{\mathbf C}_{33} &=& 1,                                                                  \nonumber \\
{\mathbf C}_{44} &=& \frac{1 + (\chi -1)\cos(\chi \tau)}{\chi}
\end{eqnarray}
and 

\begin{eqnarray}
{\mathbf S}_{11} &=& 
\frac{\omega}{\varsigma_m} \frac{ \pmb{O}_{ \! m \alpha }^2 \sin(\chi\tau)\, + \, 
        \chi \pmb{O}_{ \! m \beta}^2  \sin\tau         } {  ( \chi - 1)\chi},            \nonumber \\
{\mathbf S}_{12} &=& {\mathbf S}_{21} = \frac{\omega}{\varsigma_m}
\frac{ \pmb{O}_{ \! m \alpha }\pmb{O}_{ \! m \beta} 
           [   \sin(\chi\tau) -\chi \sin\tau ]  }
      {  ( \chi - 1 ) \chi} ,                                                            \nonumber \\
{\mathbf S}_{13} &=& {\mathbf S}_{31} = {\mathbf S}_{23} = {\mathbf S}_{32} = 
{\mathbf S}_{33} = {\mathbf S}_{34} = {\mathbf S}_{43} = 0,                              \nonumber \\
{\mathbf S}_{14} &=& {\mathbf S}_{41} = 
-\sqrt{\frac{\omega}{\varsigma_m}}
\frac{\pmb{O}_{ \! m \alpha }}{\chi}\sin( \chi \tau),                                   \nonumber \\
{\mathbf S}_{22} &=& 
\frac{\omega}{\varsigma_m}\frac{   \pmb{O}_{ \! m \alpha }^2 \chi \sin\tau \, + \, 
          \pmb{O}_{ \! m \beta}^2 \sin(\chi\tau)          }
      {  ( \chi - 1 ) \chi},                                                             \nonumber \\
{\mathbf S}_{24} &=&  {\mathbf S}_{42} = 
-\sqrt{\frac{\omega}{\varsigma_m}}\frac{ \pmb{O}_{ \! m \beta}}{\chi} \sin(\chi\tau),  \nonumber \\       
{\mathbf S}_{44} &=& \frac{\chi - 1 }{\chi} \sin(\chi\tau),                             
\end{eqnarray}
where we have defined 
\begin{equation}
\tau := \epsilon t/4,  \,\,\, \chi := 
\frac{\omega}{\varsigma_m} \pmb{O}_{ \! m \alpha }^2 + 
\frac{\omega}{\varsigma_m} \pmb{O}_{ \! m \beta}^2 + 1.  
\end{equation}

The functions appearing in Eqs.~(\ref{funct}) and (\ref{funct2}) are
\begin{eqnarray}
F &:=&  
\frac{\omega^2}{\varsigma_m^2}
\frac{\pmb{O}_{ \! m \alpha }^2 {\pmb O}_{m \beta}^2}{\chi(\chi - 1)} 
\left[ \tfrac{\chi - \cos[(\chi - 1)\tau]}{(\chi-1)} + 
\tfrac{ \cos (\chi\tau) - 1}{\chi}  
-{ \cos \tau }  \right],          \nonumber \\
G &:=&
\frac{\omega^2}{\varsigma_m^2}\frac{ \pmb O_{m \alpha }^4}{(\chi-1)^2}  + 
\frac{\omega^2}{\varsigma_m^2} 
\frac{\pmb O_{m \beta}^4 (\chi^2 -2\chi + 2)}{\chi^{2}(\chi-1)^{2} },                    \nonumber \\ 
H &:=&  
\frac{\omega}{\varsigma_m}\frac{\pmb O_{m \beta }^2}{\chi^{2}} + 
\frac{\omega^2}{\varsigma_m^2}
\frac{\pmb O_{m \alpha}^2\pmb O_{m \beta}^2(\chi^2 -\chi + 1)}
      {\chi^{2}(\chi-1)^{2} },                                                           \nonumber \\
I &:=& \frac{\omega^2}{\varsigma_m^2}\frac{{\pmb O}_{m \alpha}^2 {\pmb O}_{m \beta}^2}
                                            {\chi(\chi - 1)} \!
\left[ \tfrac{\cos[(\chi - 1)\tau]}{(\chi-1)} + 
       \frac{ {\pmb O}_{m \beta}^2}{{\pmb O}_{m \alpha}^2} \tfrac{\cos[\chi\tau] }{ \chi} +
       { \cos \tau }   
\right].                                                                                 \nonumber \\
                                                                                         \label{auxfunc}
\end{eqnarray}


\end{document}